\documentclass[
    twocolumn, superscriptaddress, nofootinbib, amsmath, amssymb,aps, floatfix
]{revtex4-1}
\usepackage{xr}
\usepackage[usenames,dvipsnames,table]{xcolor}
\usepackage{graphicx,epsfig,psfrag,bm,amssymb}
\usepackage[colorlinks=true,urlcolor=blue
,anchorcolor=blue,citecolor=blue,filecolor=blue,linkcolor=blue,menucolor=blue
]{hyperref}
\usepackage{dcolumn}
\usepackage{bm}
\usepackage{color}
\usepackage{soul}
\usepackage{cancel}
\usepackage{mathrsfs,amsfonts,color}
\usepackage{caption}
\usepackage{subcaption}
\usepackage{comment}
\usepackage{xspace}
\usepackage[capitalise]{cleveref}
\usepackage{ulem}
\usepackage{cancel}

\newcommand{\beq}{\begin{equation}}
\newcommand{\eeq}{\end{equation}}

\begin{document}

\title{Focusing the Axion Wind with Ferrite Flux Concentrators}
\author{Christina Gao}
\email{gaoy3@sustech.edu.cn}
\affiliation{Department of Physics, Southern University of Science and Technology, Shenzhen, China}
\author{Yuxuan He}
\email{he.yx25@cityu.edu.hk}
\affiliation{Department of Physics, City University of Hong Kong, Kowloon, Hong Kong SAR, China}
\author{Yubo Mo}
\email{12210407@mail.sustech.edu.cn}
\affiliation{Department of Physics, Southern University of Science and Technology, Shenzhen,  China}
\author{Peiran Yin}
\email{ypr@nju.edu.cn}
\affiliation{National Laboratory of Solid State Microstructures and Department of Physics, Nanjing University, Nanjing, China.}

\date\today

\begin{abstract}
Axion dark matter may couple to fermion spins through an effective oscillating magnetic field, the ``axion wind". Existing approaches for detecting axion wind via axion-electron coupling $g_{ae}$ face sensitivity limitations due to the extremely weak pseudomagnetic field generated, typically a few times $10^{-18}$ tesla for $g_{ae} \sim  10^{-10}$. We propose to use ferrimagnetic flux concentrators to amplify the pseudomagnetic field into a genuine magnetic field, which can then be measured using general precision magnetometry techniques. 
This greatly expands the experimental possibilities beyond conventional methods that rely on detecting small transverse magnetic fields from spin precession induced by axion wind.
Here we utilize the libration mode of a levitated ferromagnet, which can achieve sub-femtotesla sensitivity, enabling detection of axion-electron coupling strengths $g_{ae} \lesssim 10^{-11}$ in the frequency range $1-500$ Hz.
\end{abstract}

\maketitle
\flushbottom

\label{sec:intro}
Axions~\cite{Peccei:1977hh,Weinberg:1977ma,Wilczek:1977pj} and axion-like particles (ALPs)~\cite{Svrcek_2006} arise in well-motivated extensions of the Standard Model (SM) of particle physics.
Due to their pseudo-Goldstone boson nature, 
axions and ALPs are naturally light and interact feebly with ordinary matter. This ensures their cosmological stability, making them viable dark matter (DM) candidates~\cite{Abbott:1982af, Preskill:1982cy, Dine:1982ah}. Their masses could span a vast range from $10^{-22}$ eV to a few eVs. If they make up all the DM, given a local density of $\rho_{\rm DM}\sim 0.4$ GeV/cm$^3$~\cite{de_Salas_2021}, the occupation number per de Broglie volume $\gg1$ for axions and ALPs. Therefore, the axion DM behaves like wave rather than particle, and can be treated as a classical oscillating field with frequency set by its mass $m_a$.

Axion wind experiments~\cite{Abel:2017rtm,JacksonKimball:2017elr,Wu_2019,Garcon:2019inh,Bloch:2019lcy, Bloch:2021vnn, Jiang:2021dby, Bloch:2022kjm, Abel:2022vfg,Gao:2022nuq,Lee:2022vvb, Wei:2023rzs,Xu:2023vfn, Gavilan-Martin:2024nlo} represent an innovative frontier in searches for axion DM.  
These experiments leverage the effective interaction between axions and the SM fermions, which in the non-relativistic limit is dominated by the term  
\begin{equation}\label{eq:V}
V= -\gamma_f {\bf H}_a\cdot {\bf S},
\end{equation}
where ${\bf S}$ is the fermion spin, $ \gamma_f$ is the gyromagnetic ratio. For electrons, $\gamma_e=-g_e \frac{\mu_B}{\hbar}$. The pseudomagnetic field generated by the axion-electron coupling takes the form of 
\begin{equation}\label{eq:Ha}
{\bf H}_a \simeq \frac{g_{ae}}{e}\sqrt{2\rho_{\rm DM}} \cos(m_at) {\bf v}_a,
\end{equation}
where $g_{ae}\equiv \frac{C_e m_e}{f_a}$ is the dimensionless axion-electron coupling, whose size is related to the new global symmetry breaking scale $f_a$ and a model-dependent coefficient $C_e$~\cite{Rosenberg2024Axions}. $|{\bf v}_a|=v\sim 10^{-3}c$~\cite{Evans:2018bqy} is the typical DM velocity in our galaxy. 
Thus ${\bf H}_a$ has a typical magnitude of 
\beq\label{eq:Ha_size}
|{\bf H}_a|\sim 4\times 10^{-18}{\rm T} \left(\frac{g_{ae}}{10^{-10}}\right)~.
\eeq 

For axions below a few eVs, the most severe constraints on axion-electron coupling $g_{ae}$ come from considering axions sourced by astrophysical objects, including star cooling limits from red giants~\cite{Capozzi:2020cbu} and direct searches for solar axions~\cite{XENON:2022ltv,Gondolo:2008dd,Arisaka:2012pb}. 
Laboratory probes and haloscopes that depend little on astrophysical environments can be complementary to the searches above, even though they have not reached the same level of sensitivity~\cite{Bloch:2019lcy,Yan:2019dar,Terrano:2015sna,Terrano:2019clh, Brandenstein:2022eif,Alonso:2018dxy,Crescini:2018qrz, QUAX:2020adt,terrano2015shortrangespindependentinteractionselectrons,Ikeda_2022,Yan_2019}.

For $m_a/(2\pi)\gg 1$ kHz, several experimental proposals have been put forward to search for $g_{ae}$, including approaches based on RF cavities~\cite{Flower:2018qgb,Chigusa:2020gfs}, quantum sensors~\cite{Ikeda:2021mlv,Chu:2019iry} and solid-state systems~\cite{Sikivie:2014lha,Hochberg:2016sqx,Hochberg:2016ajh,Arvanitaki:2017nhi,Arvanitaki:2021wjk,Chen:2022pyd,Chigusa:2023hmz,Mitridate:2023izi,Berlin:2023ubt}. 
For $m_a/(2\pi)\leq 1$ kHz, recent developments in quantum sensing have enabled new approaches to probe $g_{ae}$, including NV centers~\cite{Chigusa:2023roq} and levitated magnetometers~\cite{Kilian:2024fsg,Ahrens2025,Higgins:2023gwq,Kalia:2024eml}, most of which propose to detect the spin precession generated by \cref{eq:V}.

In this work we propose to use ferrimagnetic flux concentrators to turn the pseudomagnetic field ${\bf H}_a$ into a \emph{genuine magnetic field} that could be order of magnitude larger than $|{\bf H}_a|$, which can then be measured using general precision magnetometry techniques. This is first pointed out by Ref.~\cite{JacksonKimball:2016wzv}, but only qualitatively. 
Here we choose to utilize the libration mode of a levitated ferromagnet, which can achieve sub-femtotesla sensitivity~\cite{JacksonKimball2016,Vinante2021,Ahrens2025,Ji:2025yvn}. 
Throughout this work, we use natural units, i.e., $\hbar = c = \varepsilon_0 = \mu_0 = 1$.

\section*{Axion induced Magnetization}

In paramagnetism and diamagnetism, the magnetization is linearly related to the magnetic field $\bf H$. 
For ferromagnetism or ferrimagnetism, this linear relation no longer holds. Instead, magnetization could depend on the history of the magnetic field $\bf H$. The magnetic flux density or magnetic induction can be written as
\beq\label{eq:BMH}
{\bf B}= {\bf M}({\bf H})+{\bf H}+{\bf M}_r~,
\eeq
where we split the magnetization into two parts. 
${\bf M}_r$ is the remanent magnetization that was there due to the history of the applied field. $\bf M(\bf H)$ can depend on $\bf H$ in a complicated way, and its dependence is usually presented using a hysteresis loop. In our treatment of magnetization, the total magnetization is 
${\bf M}({\bf H})+{\bf M}_r$. 
From \cref{eq:V,eq:Ha} the axion wind introduces a weak oscillating pseudomagnetic field to the ferromagnet\footnote{Here we ignore the orbital angular momentum $L$, which is a good approximation for 3d ions since $L$ is quenched for most 3d ions, including iron ions~\cite{blundell2001magnetism}.}. Since ${\bf H}_a$ only exists inside the material and depends on its gyromagnetic ratio, the axion induced magnetization behaves as if it were a remanent magnetization.

To determine the effect of ${\bf H}_a$ quantitatively, we treat it as a perturbation. 
Let the ferromagnet have an internal field ${\bf H}_{\rm in}$ with $|{\bf H}_{\rm in}| \gg|{\bf H}_a|$ and a total magnetization ${\bf M}({\bf H}_{\rm in})+{\bf M}_r$. 
A change in remanent magnetization due to the axion will perturb the internal field ${\bf H}_{\rm in}$ and thus the part of magnetization that depends on it. For simplicity, assume that all the fields including ${ H}_a$ are directed along the same axis. To leading order,
$
 M(H_{\rm in}+\delta H_{\rm in})\approx M(H_{\rm in})+\delta H_{\rm in} M'(H_{\rm in})$, and therefore, 
\beq\label{eq:linearise}
 \delta M\approx \chi_{\rm eff}\delta H_{\rm in}, ~{\rm with}~\chi_{\rm eff}\equiv M'(H_{\rm in})~.
\eeq
Since $ |\delta H_{\rm in}| \ll| H_{\rm in}|$ is a small perturbation along the hysteresis curve, it is a good approximation that $\chi_{\rm eff}$ remains constant in the field range $ H_{\rm in}\pm\delta H_{\rm in}$. 
The change in flux density inside the ferromagnet is thus approximately given by
\beq\label{eq:model_B}
\delta{ B}_{\rm in}=\mu_{\rm eff} \delta { H}_{\rm in}+\chi_{\rm eff}{ H}_a,~{\rm with}~\mu_{\rm eff}=\chi_{\rm eff}+1.
\eeq
Therefore, $\chi_{\rm eff}{ H}_a$ can be identified as the remanent magnetization caused by the axion wind, and the change in the total magnetization is
\beq\label{eq:model_M}
\delta { M}^{\rm tot}=\chi_{\rm eff}(\delta {H}_{\rm in}+{ H}_a)~,
\eeq
which is linearly related to the field. Thus, the ferromagnet shows the behavior of paramagnetism under the perturbation of a weak field~\cite{JacksonKimball:2016wzv}. 
Equations~(\ref{eq:model_B}) and~(\ref{eq:model_M}) give our baseline model in modelling the effect of axion wind inside the ferromagnet. Outside the material, there is no more magnetization, and we have 
\beq\label{eq:model_out}
\delta {\bf B}_{\rm out}=\delta {\bf H}_{\rm out}~.
\eeq

\subsubsection*{Ferromagnet of Arbitrary Shape}

For general shapes, we need to solve the corresponding magnetostatic problem in light of \cref{eq:model_B,eq:model_out}. 
This is outlined in \cref{app:demag} for general ferromagnetic problems. 
From now on, we drop $\delta$'s everywhere, keeping in mind that it is the axion induced field that is of interest here. In the absence of free currents,
\beq\label{eq:maxwell}
\nabla\times {\bf H}=\nabla\cdot {\bf B}=0,
\eeq
The general solution is given by 
\beq\label{eq:phi}
{\bf H}= -\nabla \Phi
\eeq 
where $\Phi$ is a scalar potential. Using \cref{eq:model_B,eq:phi,eq:maxwell}, we have
\beq\label{eq:phi_ina}
\nabla^2 \Phi_{\rm in}=  \frac{\chi_{\rm eff}}{1+\chi_{\rm eff}}\nabla\cdot{\bf H}_a~.
\eeq 
The boundary conditions are that the tangential component of $\bf H$ and the normal component of $\bf B$ need to be continuous, thus at the surface of the ferromagnet 
\beq\label{eq:phi_bca}
\Phi_{\rm in}=\Phi_{\rm out},~\mu_{\rm eff}\frac{\partial \Phi_{\rm in}}{\partial n}-\frac{\partial \Phi_{\rm out}}{\partial n}= \chi_{\rm eff}{\bf H}_a\cdot{\bf n}~.
\eeq
where  $\bf n$ is the unit normal vector pointing outward from the ferromagnetic material to air. 

\subsubsection*{Shielding of the Axion Wind}\label{app:shielding}

As an example, we solve \cref{eq:phi_ina} for the magnetostatic problem of magnetic shielding made from two concentric spherical shells with inner radius $a$ and outer radius $b$. The shielding material is assumed to have a high relative permeability $\mu_{\rm eff}$. 
Let ${\bf H}_a=h_a \hat z$. 
Due to the azimuthal symmetry, the magnetic potential can be written as
\begin{equation}
\Phi(r,\theta) = \sum_{l=0}
\left\{
\begin{array}{lc}
\frac{\alpha_l}{r^{l+1}} P_l(\cos \theta)& \quad r > b, \\
\left(\beta_l r^l + \frac{\gamma_l}{r^{l+1}}\right)P_l(\cos \theta) &\quad b > r > a, \\
\delta_l r^l P_l(\cos \theta) &\quad r < a.
\end{array}   \right . 
\end{equation}
Given ${\bf H}_a$ is constant in $z$, only $l=1$ terms can contribute. Imposing \cref{eq:phi_bca},  
we get 
\begin{align}
\alpha_1 &= \chi_{\rm eff} h_a (2\mu_{\rm eff}+1)(a^3 - b^3)b^3f^{-1}; \\
\beta_1 &= \chi_ah_a\left(2\chi_{\rm eff} a^3 - (2\mu_{\rm eff}+1)b^3\right)f^{-1}; \\
\gamma_1 &= \chi_{\rm eff} h_a 3a^3b^3f^{-1}; \\
\delta_1 &= \chi_{\rm eff} h_a 2(a^3-b^3)\chi_{\rm eff} f^{-1}, 
\end{align}
where 
\beq
f(a,b,\mu_r)=2 a^3 (-1 + \mu_{\rm eff})^2 - b^3 (2 + 5 \mu_{\rm eff} + 2 \mu_{\rm eff}^2)~.
\eeq
Using \cref{eq:phi}, the magnetic field in the cavity enclosed by the shield is approximately
\begin{equation}
\mathbf{H} (r<a) = -\mathbf{H}_a \chi_{\rm eff}\left(\frac1{\mu_{\rm eff}} + \mathcal{O}(\mu_{\rm eff} ^{-2})\right) .
\end{equation} 
Since $\mu_{\rm eff}\sim \chi_{\rm eff}\gg1$, the induced magnetic field is of the same order as that of the effective field from the axion wind and needs to be taken into account when designing low frequency axion wind detection experiments. 

\subsubsection*{Ellipsoidal Ferromagnet}

We can estimate of the size of magnetization caused by the axion wind in the simple case of an ellipsoidal ferromagnet. From \cref{app:demag},
$
 {\bf H}_{\rm in}=-N_m {\bf M}^{\rm tot}$, 
where $N_m$ is the demagnetizing factor which only depends on the geometry of the ellipsoid. Using \cref{eq:model_M}, we obtain
\beq\label{eq:Ma}
{\bf M}^{\rm tot}_{\rm ellipsoid}=\frac{\chi_{\rm eff}}{1+\chi_{\rm eff}N_m} {\bf H}_a~.
\eeq
Given $N_m$ is of order $10^{-1}$ (c.f. \cref{app:demag}), in a highly permeable material where $\chi_{\rm eff}\to\infty$, the axion induced magnetization is approximately given by $N_m^{-1}{\bf H}_a$. This suggests that we may detect an enhanced axion wind signal by placing the detector close to the surface of the ferromagnetic body. Furthermore, we can design the geometry of the ferromagnet that gives a small $N_m$ along one of the principal axes, e.g. the $z-$axis in a thin rod. 

\section*{Proposed Setup}
\begin{figure}[t]
    \centering
\includegraphics[width=0.83\linewidth]{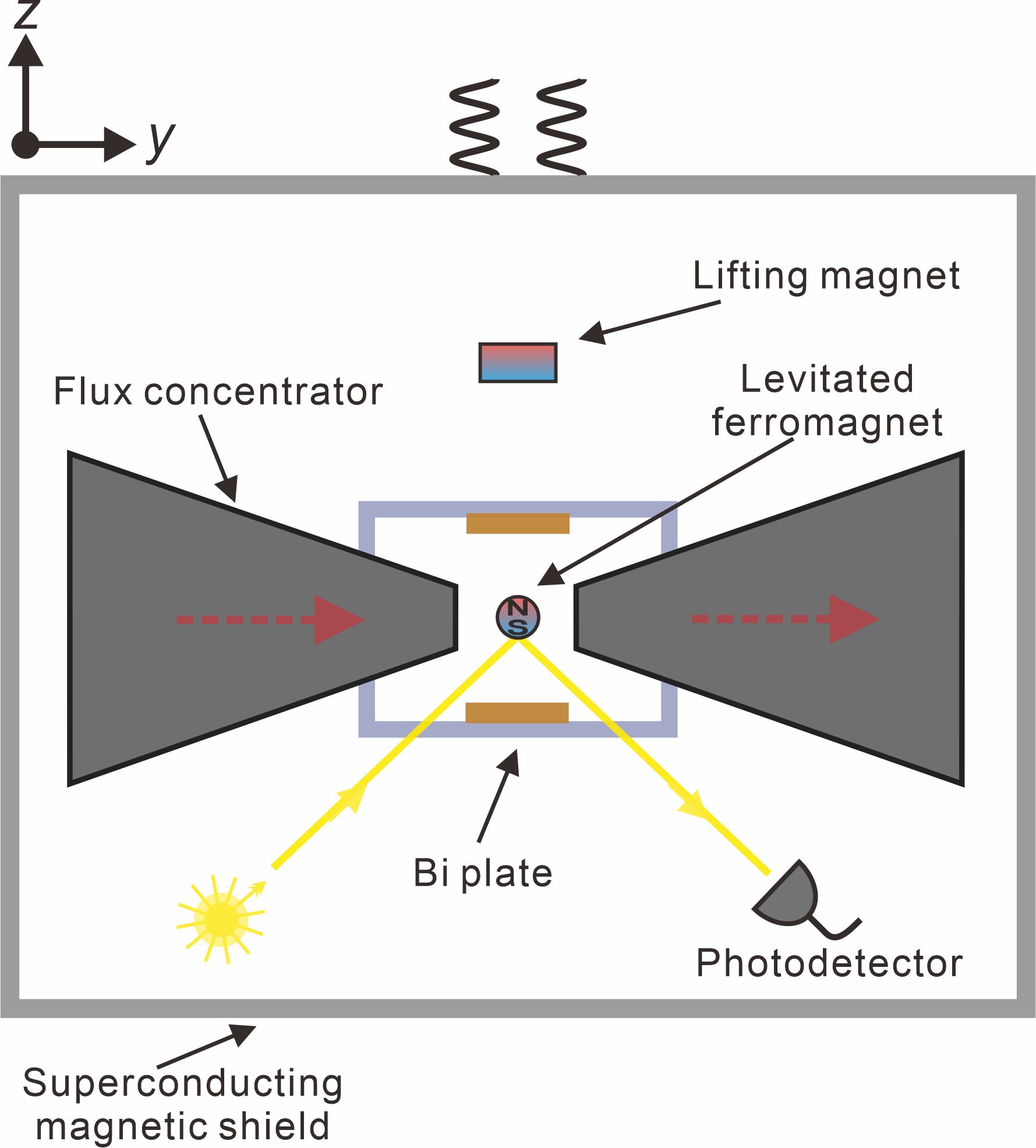}
    \caption{Schematic of the proposed axion wind detection experiment. A hard ferromagnet is magnetically levitated at the center of the apparatus using permanent lifting magnets and a diamagnetic Bi plate, with the dipole moment lying along the $z$-axis.  Ferrimagnetic flux concentrators are positioned strategically around the levitated ferromagnet to amplify the weak pseudomagnetic field ${\bf H}_a$ generated by axion-electron interactions. The rotational motion of the dipole is monitored using an optical method. The entire setup is enclosed within a superconducting magnetic shield to suppress ambient magnetic noise while preserving the coherent axion-induced signal.}
    \label{fig:setup}
\end{figure}

After establishing the effect of axion wind and how to model it in ferromagnetism, we propose to look for ${\bf H}_a$ using a levitated ferromagnet. 
The proposed experimental setup, illustrated in \cref{fig:setup}, consists of a flux concentrator and a levitated ferromagnetic magnetometer.

\begin{figure}[t]
    \centering
\includegraphics[width=1\linewidth]{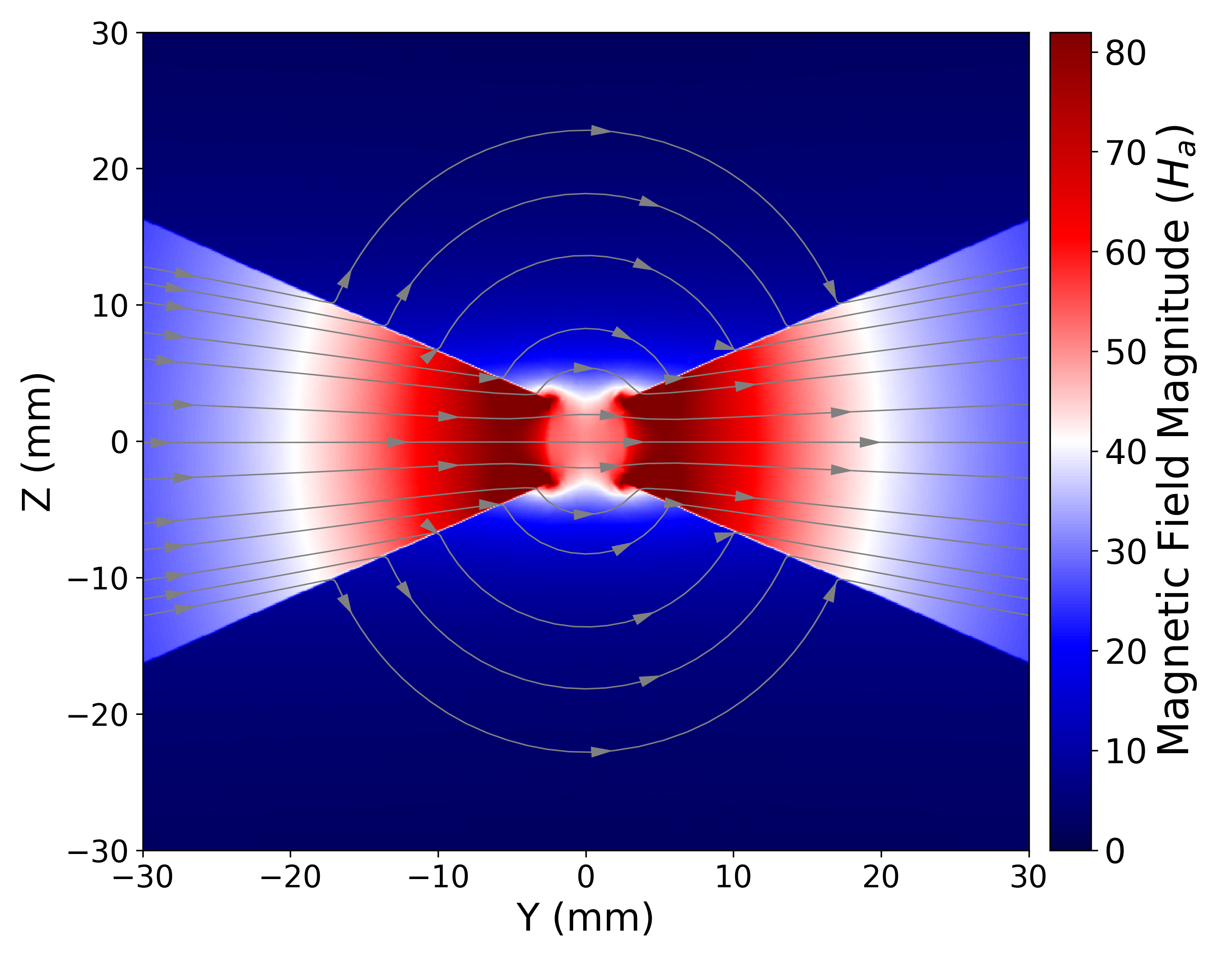}
    \caption{Numerical simulation of the magnetic field in the $y-z$ plane assuming a residual magnetization $\chi_{\rm eff} {\bf H}_a$ for the inner 60 mm$\times$60 mm region of the setup. The two cone-like flux concentrators are separated by 5 mm from tip to tip, and each has a tip radius of 3 mm, a base radius of 100 mm, and a height of 200 mm. }
    \label{fig:field}
\end{figure}

The key design is the ferrimagnetic flux concentrator positioned strategically around the levitated sensor. These concentrators are designed to amplify the weak pseudomagnetic field ${\bf H}_a$ through the mechanism approximately described by \cref{eq:Ma}. To obtain the field configuration in between the concentrators, one can solve the corresponding magnetostatic problem assuming a residual magnetization provided by the axion wind $\chi_{\rm eff} {\bf H}_a$. This is done numerically in \cref{app:signal_simulation}. Figure~(\ref{fig:field}) shows the resultant magnetic field in the $y-z$ plane in the inner 60 mm$\times$60 mm region of the setup. Between the tips of the two concentrators, the field is nearly uniform and has a magnitude of approximately 50 times $|{\bf H}_a|$. Therefore, the specific design in \cref{fig:field} amplifies the axion signal by 50 times.

At the center of the setup is a levitated hard ferromagnet serving as the sensor. The ferromagnet with dipole moment lying along the $z$-axis, is magnetically levitated by lifting magnets and diamagnetic bismuth (Bi) plates. The trap potential would be adjusted by the ferrimagnetic flux concentrators. To achieve stable diamagnetic levitation, the bismuth plate needs to be placed close to the levitated magnet. In the regime that the rotational angular momentum dominates over intrinsic spin, the ferromagnet librates about the background magnetic field, with the gyroscopic dynamics negligible \cite{JacksonKimball2016,Vinante2021,Ahrens2025,Ji:2025yvn}. In our setup, we utilize the libration mode around the $x-$axis to detect the magnetic field in the $y$ direction.  
The rotational motion of the levitated ferromagnet is monitored by collecting the reflected light. A laser beam is directed at the ferromagnet, and the reflected light is captured by a photodetector. Small angular displacements $\theta(t)$ of the levitated ferromagnet, induced by the time-varying magnetic torque from ${\bf H}_a$, are converted into measurable optical intensity variations.

To suppress the ambient magnetic noise, the entire apparatus is enclosed within a superconducting magnetic shield. The choice of the shielding is critical, since shields made of a soft ferro- or ferrimagnetic material could suppress the axion wind signal~\cite{JacksonKimball:2016wzv}. In order not to ``block" the axion wind, we propose to use a superconducting magnetic shield. This would require a cryogenic environment, which also helps suppress the thermal noise.

\subsubsection*{Noise and Sensitivity} 

Let the ferromagnet have a dipole moment ${\bm m}$, and a resonant rotational frequency $\omega_0$ around the $x-$axis with a corresponding moment of inertia $I$. Its angular displacement about the $x-$axis obeys the following equation of motion
\beq\label{eq:eom1}
I \ddot{\theta}+I\gamma \dot{\theta}+I\omega_0^2 \theta=({\bm m}\times {\bf B}(t))\cdot \hat x+\tau_{\rm noise},
\eeq
where $\gamma$ ($\ll \omega_0$) characterizes the mechanical dissipation, and $\tau_{\rm noise}$ represents noises including thermal noise and backaction during optical measurements. Here, the ambient magnetic noise is neglected as it should be effectively shielded by the superconducting enclosure. We also assume that vibrational noise is sufficiently suppressed above 1 Hz. 
The maximum signal torque is $|{\bm m}|B$, where $B$ is the amplitude of the magnetic field.
In Fourier space, \cref{eq:eom1} can be written as
\beq\label{eq:eom2}
\tilde{\theta}(\omega)=\frac{|{\bm m}| \tilde B(\omega)}{I(\omega^2-\omega_0^2+i\gamma\omega)}= \frac {|{\bm m}|} {I}\chi(\omega)\tilde B(\omega),
\eeq
where $\theta(t)=\int_{-\infty}^\infty d\omega/(2\pi) e^{i\omega t} \tilde \theta(\omega)$, and
\beq
\chi(\omega,\omega_0)=\frac1{\omega^2-\omega_0^2+i\gamma\omega}~
\eeq
is the mechanical response of the oscillator. From \cref{eq:eom2}, we obtain a relation between the power spectral density (PSD) of $\theta$, defined as $S_{\theta\theta}(\omega)=\frac {|\tilde \theta(\omega)|^2}{t_{\rm int}}$, and the PSD of the magnetic field $S_{BB}(\omega)$: 
\beq\label{eq:SBB}
S_{BB}(\omega)=\frac{I^2}{|{\bm m}|^2|\chi(\omega)|^2}S_{\theta\theta}(\omega). 
\eeq
If the ferromagnetic sensor  experiences several sources of magnetic fields, their contributions to $S_{\theta\theta}$ simply add up due to the linear nature of \cref{eq:eom1}.

The axion induced magnetic field takes the form of $B_a \cos(\omega_a t)$,  where $\omega_a=m_a(1+\frac12v^2/c^2)$. Given $v/c\sim10^{-3}$, $B_a$ has a width roughly given by $\gamma=\frac12 m_av^2/c^2\sim 10^{-6}m_a$. 
Thus, the axion PSD is $
S^{\rm axion}_{BB}(\omega)= B_a^2 \frac{\pi}2(\delta(\omega-\omega_a)+\delta(\omega+\omega_a))
$.  
In the limit of long integration time\footnote{Here, we use the following relation: $\pi \delta(x)=\lim_{t\to\infty} t~{\rm sinc}(x t)$.} $t_{\rm int}$, 
\beq\label{eq:Ssig}
\begin{split}
S^{\rm sig}_{BB}(\omega)&\approx B_a^2 t_{\rm tot} 
{\rm sinc}\left((\omega-\omega_a) t_{\rm int}\right),\\
t_{\rm tot}&=\left\{
\begin{array}{cc}
t_{\rm int} & t_{\rm int}\leq\tau_a,\\
\sqrt{t_{\rm int}\tau_a} &t_{\rm int}>\tau_a,
\end{array}
\right .
\end{split}
\eeq
where $\tau_a\sim 2\pi/(10^{-6}m_a)\sim 10^4~{\rm sec}\left(\frac{100~\rm Hz}{m_a/(2\pi)}\right)$  is the coherence time of the axion wind.  
For the axion masses considered in this work, $\gamma_a\ll \gamma $ holds for most of them. Hence, from now on we assume $\omega_a\simeq m_a$\footnote{A full treatment taking into account of the finite width of axion wind can be found in a companion paper.}.

Apart from the axion wind, the levitated sensor also experiences thermal noises, including Brownian thermal noise and the magnetic hysteresis noise from ferrimagnetic flux concentrators. The PSD of Brownian thermal torque noise is given by $4I\gamma k_B T_{\rm en}$~\cite{Yin:2022geb, Yin2025}, where $T_{\rm en}$ is the environment  temperature. The corresponding PSD in $B$ is thus given by
\beq\label{eq:Sth}
S_{BB}^{\rm th}(\omega)=\frac{4I\gamma k_B T_{\rm en}}{|{\bm m}|^2} ~.
\eeq

The measurement imprecision $S^{\rm imp}_{\theta\theta}$ and backaction $S^{\rm ba}_{\tau \tau}$ satisfy the relation 
\beq\label{eq:uncertainty}
\sqrt{S^{\rm imp}_{\theta\theta}S^{\rm ba}_{\tau \tau}}=  \hbar/(2\sqrt{\eta_{\rm d}}),
\eeq
with $\eta_{\rm d}$ being the detection efficiency~\cite{Clerk2010}. Detection attains the standard quantum limit (SQL) at a unit detection efficiency. Here we choose to detect the rotation of the levitated ferromagnet with optical methods, which has attained $\eta_{\rm d}\sim 0.01$ with current experimental techniques. The total angle noise during the magnetic field measurement can be expressed as:
\begin{equation}
\begin{split}
S&^{\rm noise}_{\theta\theta}= S^{\rm th}_{\theta\theta}+S^{\rm imp}_{\theta\theta}+S^{\rm ba}_{\theta\theta}\\
&= S^{\rm th}_{\theta\theta}+(S^{\rm imp}_{\theta\theta}+\frac{S^{\rm ba}_{\tau\tau}|\chi|^2}{I^2})\\
  &=S^{\rm th}_{\theta\theta}+(S^{\rm imp}_{\theta\theta}+\frac{\hbar^2|\chi|^2}{4\eta_{\rm d}S^{\rm imp}_{\theta\theta}I^2})\geq S^{\rm th}_{\theta\theta}+\frac{\hbar |\chi|}{I\sqrt{\eta_{\rm d}}},
\end{split}
\end{equation}
where we used $S^{\rm ba}_{\theta\theta}=S^{\rm ba}_{\tau\tau}|\chi|^2/I^2$ and \cref{eq:uncertainty} in the second and third line respectively. 
The total measurement noise is thus bounded from below with $S^{\rm mea}_{\theta\theta}\geq \hbar|\chi(\omega,\omega_0)|/(I\sqrt{\eta_{\rm d}}) $, with the minimum reached at 
$S^{\rm imp}_{\theta\theta}\big|_{\rm min}=S_{\theta\theta}^{\rm ba}\big|_{\rm min}=\frac{\hbar|\chi(\omega, \omega_{\rm 0})|}{2I\sqrt{\eta_{\rm d}}}$. This requires the detection laser power to be set to an optimal value $P^{\rm opt}(\omega)$ at each frequency~\cite{Li2023}. 

\begin{figure}[t]
    \centering
\includegraphics[width=1\linewidth]{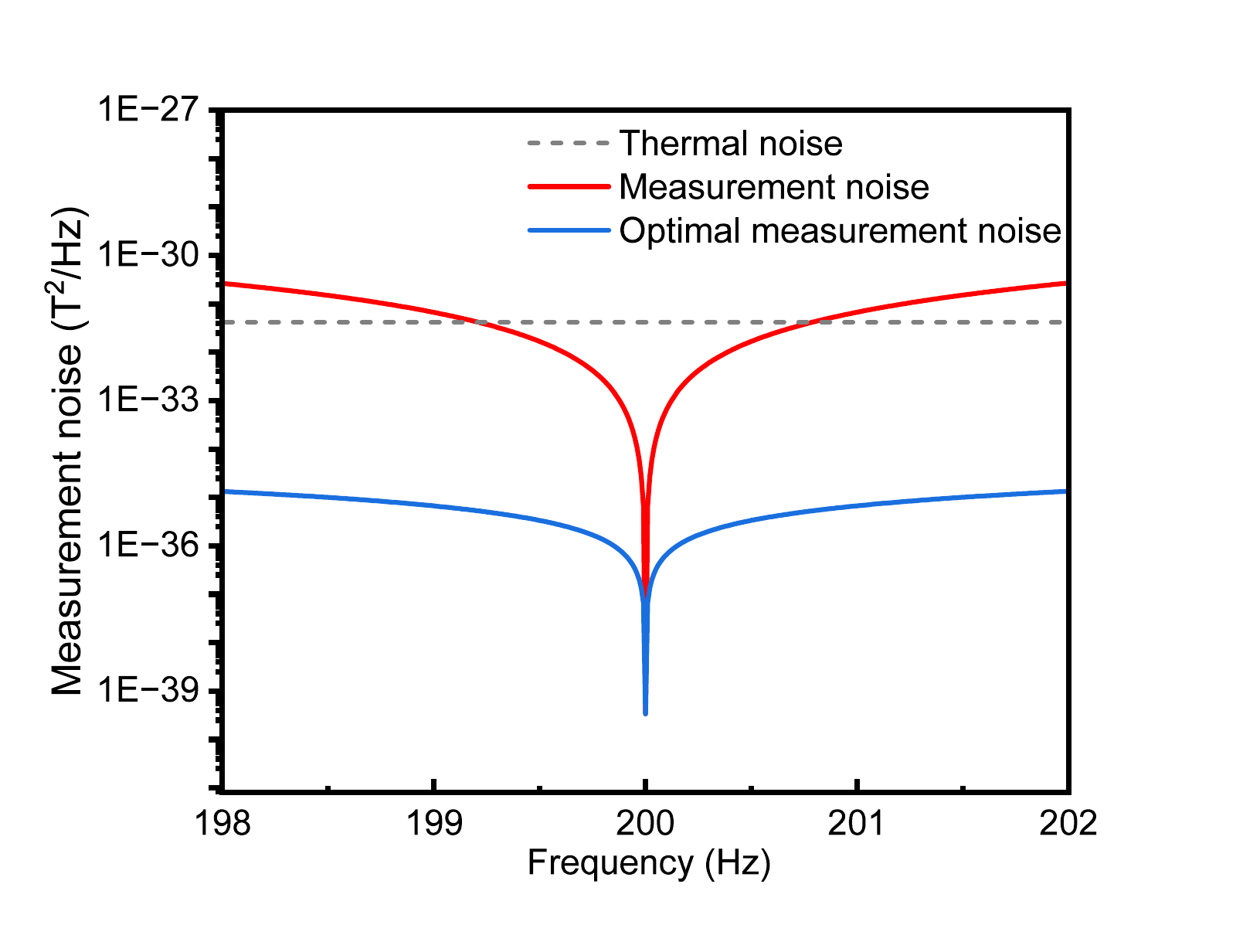}
    \caption{Comparison of measurement noise \cref{eq:S_mea}
    and thermal noise \cref{eq:Sth}. Optimal measurement noise indicates the minimum measurement noise $\frac{\hbar |\chi|}{I\sqrt{\eta_{\rm d}}}$ with $\eta_{\rm d}=0.01$.}
    \label{fig:3}
\end{figure}

In practical measurements with the resonant frequency of levitated ferromagnet fixed at $\omega_{\rm 0}$, the measurement imprecision is set to be $S^{\rm imp}_{\theta\theta}(\omega_0)=\frac{\hbar|\chi(\omega=\omega_{\rm 0}, \omega_{\rm 0})|}{2I\sqrt{\eta_{\rm d}}}$. 
With this, we use \cref{eq:SBB,eq:uncertainty} to express the total measurement noise as 
\beq\label{eq:S_mea}
S_{BB}^{\rm mea}(\omega)=\frac{I^2S_{\theta\theta}^{\rm imp}(\omega_0)}{|\bm m|^2|\chi(\omega,\omega_0)|^2}+ \frac{\left(\frac{\hbar}{2\sqrt{\eta_{\rm d}}}\right)^2}{|{\bm m}|^2 S_{\theta\theta}^{\rm imp}(\omega_0)}~.
\eeq
In \cref{fig:3} we plot $S_{BB}^{\rm mea}(\omega)$ and $S_{BB}^{\rm th}(\omega)$ (\cref{eq:Sth}) for $\omega_0/(2\pi)=200$~Hz. It is clear that thermal noise dominates up to a frequency range $\Delta\omega \equiv2|\omega^*-\omega_0|$, where $S_{BB}^{\rm th}(\omega^*)=S_{BB}^{\rm mea}(\omega^*)$. Therefore, during one measurement loop at a fixed $\omega_0$, $\Delta\omega$ is the frequency range that the thermal noise limits the magnetic sensitivity. For the benchmark 200 Hz, $\Delta\omega/(2\pi)\sim 2$ Hz and is much larger than the mechanical dissipation $\gamma/(2\pi)$.

\begin{figure*}[t]
    \centering
\includegraphics[width=0.7\linewidth]{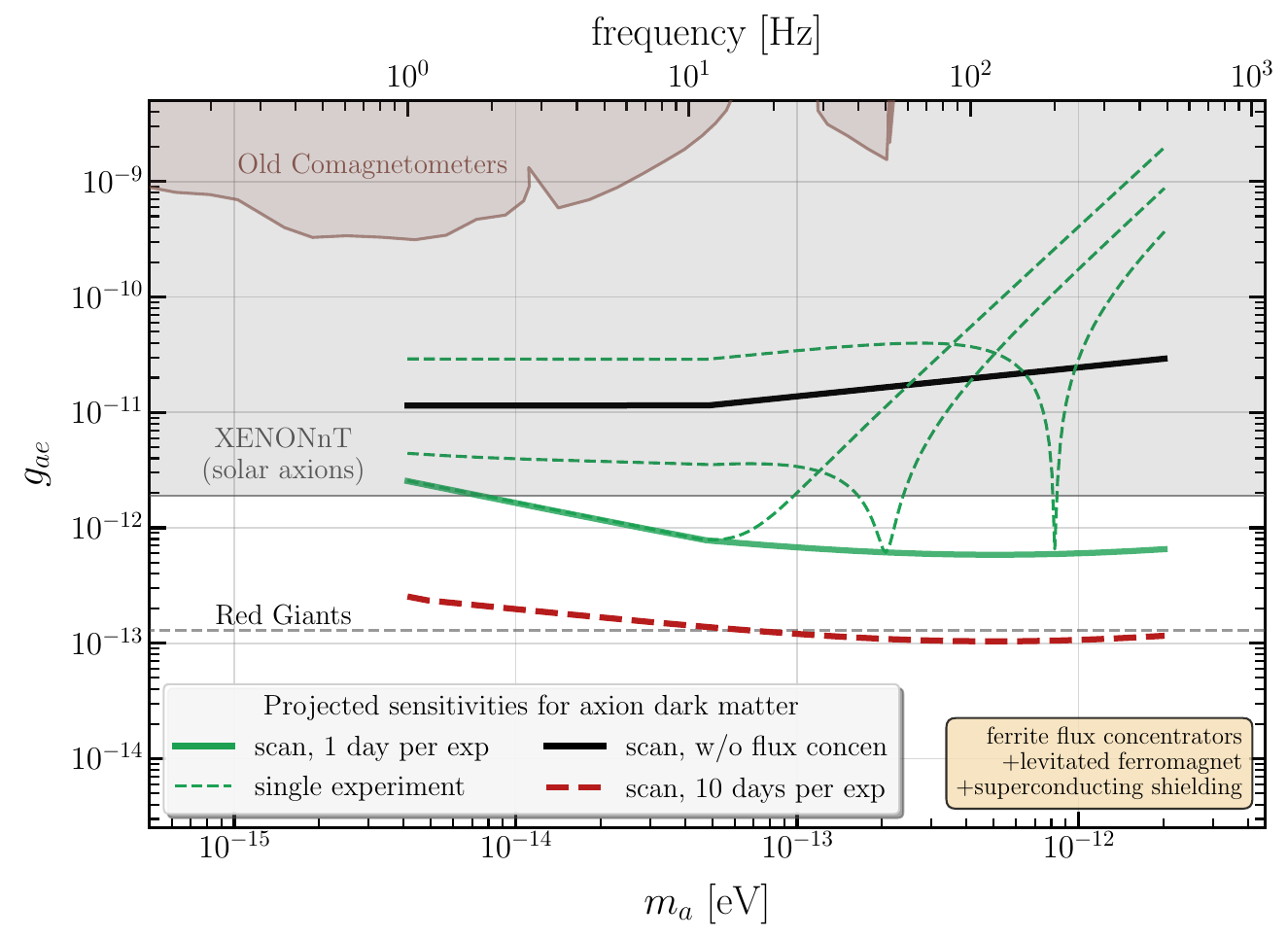}
    \caption{Projected sensitivties for axion-electron coupling in the axion mass range of 1 to 500 Hz using resonant detection of a levitated ferromagnet in a superconducting magnetic shield. The green (black) solid line assumes that each experiment has $t_{\rm int}=1$ day and a temperature of 3K with (without) a pair of ferrite flux concentrators as depicted in \cref{fig:field}. The red dashed line assumes that $t_{\rm int}=10$ days and $T_{\rm en}=0.3$K with flux concentrators. Also shown are the existing limits from comagnetometers (brown)~\cite{Bloch:2019lcy}, solar axion experiments (solid gray)~\cite{XENON:2022ltv} and star cooling argument (dashed gray)~\cite{Capozzi:2020cbu}.}
    \label{fig:proj}
\end{figure*}

Thus in the thermal dominated regime, by comparing \cref{eq:Ssig} and \cref{eq:Sth} an estimate of the setup's sensitivity to axion-electron coupling can be obtained.  
Assume that the levitated sensor is a cylinder of radius $r$ and length $l$ with density $\rho$ and remanent magnetization $M_r$\footnote{$M_r$ will receive a correction roughly equal to $2H_a$ from the previous discussion, which is negligible compared to the typical magnetization of a ferromagnet.}. In the setup depicted by \cref{fig:setup}, the moment of inertia is given by $I=\frac12\rho \pi r^4 l$. Using $|{\bm m}|= M_r \pi r^2 l$, we obtain
\begin{widetext}
\begin{equation}\label{eq:snr}
{\rm SNR}^{\rm ther~dom}=\frac{|{\bm m}|^2 B_a^2t_{\rm tot}}{4k_B \gamma T_{\rm en} I }
\sim 1 \times  \left(\frac{B_a}{5\times 10^{-19} \rm T} \right)^2 \left(\frac{l}{5\rm m m}\right)\left(\frac{M_r}{1 \rm T}\right)^2\left(\frac{8 \rm g cm^{-3}}{\rho}\right)\left(\frac{3\rm K}{T_{\rm en}}\right) \left(\frac{ 0.1 \rm m Hz}{\gamma/(2\pi)}\right) \left(\frac{t_{\rm tot}}{10^5 \rm sec}\right)~.
\end{equation}
\end{widetext}
By setting ${\rm SNR}^{\rm ther~dom}=1$ and using \cref{eq:Ha_size}, we obtain the limit on axion-electron coupling $g_{ae}\sim 10^{-11}$, which is depicted in \cref{fig:proj} as the solid black line. Since the magnetic hysteresis noise from flux concentrators is not taken into account in \cref{eq:snr}, 
this is the case when no flux concentrators are deployed and hence no amplification to the signal. 

When the flux concentrators made of soft magnetic ferrite are deployed, the magnetic noise introduced by them cannot be ignored. At low frequency ($\ll$ 100 kHz), the magnetic noise is dominated by the magnetic hysteresis losses. The levitated ferromagnet would generate a magnetic field in the flux concentrators, and the magnetic noises corresponding to hysteresis losses are determined by the complex relative permeability of the ferrite material $\mu_{\rm r}=\mu_{\rm r}'-i\mu_{\rm r}''$, and has a frequency dependence of $\omega^{-1/2}$~\cite{Kornack2007}. Here we choose to build the flux concentrators with MnZn ferrites. In our proposed setup, the background magnetic field generated by the lifting magnet and levitated ferromagnet around tip of the  flux concentrators is as low as $B_{\rm bg}\lesssim 0.05$ mT, see more details in \cref{app:filed_bkg}. At low frequency and weak magnetic field, a complex relative permeability of $\mu_{\rm r}'=3000, \mu_{\rm r}''=0.2$ is achievable. Assuming the two cone-like flux concentrators are separated by 5 mm from tip to tip, and each has a tip radius of 3 mm, a base radius of 100 mm and a height of 200 mm. We calculate the energy dissipation rate $P$ when a dipole moment $\bm m$ is oscillating at frequency $\omega$. According to the fluctuation-dissipation theorem, the magnetic noise is determined as: 
\beq\label{eq:Sfer-theory}
S_{BB}^{\rm fer}(\omega)=\frac{8 k_{\rm B} T_{\rm en} P}{\bm {|m|^2} \omega^2},
\eeq
with $P=\int_V \frac{1}{2}\omega \mu_{\rm r}''\mu_0 H^2 \rm {d} V$.
Using finite element simulations, we get
\beq\label{eq:Sfer}
S_{BB}^{\rm fer}(\omega)
\approx \frac{ \left(0.11{\rm fT}\right)^2}{\rm Hz}\left(\frac{200\rm Hz}{\omega/(2\pi)} \right)\left(\frac{T_{\rm en}}{3\rm K} \right)~.
\eeq
From \cref{app:bkg} we see that $S_{BB}^{\rm fer}(\omega)$ is almost independent of the base radius of the cone. By widening the distance between the cones, the noise term 
$S_{BB}^{\rm fer}(\omega)$ is reduced, but so is the amplification factor.  

By employing a pair of ferrite flux concentrators that brings about an amplification of signal by a factor of $N$, the sensitivity to the axion-electron coupling is approximately given by
\beq
g_{ae}\sim \frac{e\sqrt{S_{BB}^{\rm th}+S_{BB}^{\rm fer}}}{N \sqrt{2\rho_{\rm DM}t_{\rm tot}}v},~
t_{\rm tot}=\left\{
\begin{array}{cc}
t_{\rm int} & t_{\rm int}\leq\tau_a,\\
\sqrt{t_{\rm int}\tau_a} &t_{\rm int}>\tau_a.
\end{array}
\right .
\eeq
Assume that $t_{\rm int}=1$ day,  in~\cref{fig:proj} we show the projected sensitivities to the axion-electron coupling with and without flux concentrators in solid green and black lines. 
To cover the search range of $1-500$ Hz, we need to scan the resonant frequency $\omega_0$. The thermal dominated frequency range $\Delta\omega$ can thus be set as the frequency scan step, which is $\omega_0$ dependent. In \cref{fig:proj} we show three examples of single measurement loop (dashed green) at distinct resonant frequencies. To cover the entire frequency range, approximately one year of total run time is required. 
Since our setup can take data as long as 10 days, we also project the limits that can be achieved with $t_{\rm int}=10$ days in dashed red line, assuming a further suppression of thermal noise with $T_{\rm en}=0.3$ K. 
For comparion, we also show existing bounds from axion dark matter detection using comagnetometers (brown)~\cite{Bloch:2019lcy}\footnote{For old comagnetometers, the new recast Ref. \cite{Lee:2022vvb} obtained more stringent bounds on axion-neutron couplings, which can be translated to bounds on axion-electron couplings~\cite{Gavilan-Martin:2024nlo}. Here we only show the older one that is cited by \cite{AxionLimits}.}, XENONnT experiments searching for solar axions (solid gray)~\cite{XENON:2022ltv} and star cooling limits (dashed gray)~\cite{Capozzi:2020cbu}. Our proposed setup with no flux concentrators could reach a sensitivity surpassing the existing axion wind limit, and the future setups with ferrite flux concentrators and longer integration time could exceed the astrophyiscal bounds for frequencies higher than 20 Hz.

\section*{Conclusions}
In conclusion, we have demonstrated that the combination of ferrimagnetic flux concentrators and a levitated ferromagnet enables an effective search for axion dark matter via axion-electron coupling, with projected sensitivities surpassing the current experimental and astrophysical constraints. Using the flux concentrators, we are able to amplify the magnetic signals of axion-electron coupling by 50 times.  We have presented a detection strategy for the expected magnetic fields using a levitated ferromagnet—an emerging ultrasensitive magnetometry technique capable of achieving sub-femtotesla sensitivity within the frequency range of 1 to 500 Hz. Our analysis has accounted for the dominant noise sources, including Brownian thermal noise, magnetic noise, and measurement noise. We also demonstrate that conventional $\mu$-metal magnetic shielding would attenuate the axion-electron coupling signal. Therefore, we propose the use of a superconducting magnetic shield instead. Further optimizations of the proposed setup—such as 
reducing mechanical dissipation, and 
operating at lower temperatures to improve magnetic sensitivity, employing a network of separated sensors—could significantly enhance its potential for probing physics beyond the Standard Model~\cite{Higgins:2023gwq, Kalia:2024eml, Tian2025}.
\acknowledgments

We thank Dmitry Budker for many helpful comments on the manuscript.   
C.G. acknowledges the Aspen Center for
Physics for its hospitality where part of this work is done, which is supported by National Science Foundation grant PHY-2210452. Y.H. is supported by the GRF grants No. 11302824 and No. 11310925 from the Hong Kong Research Grants Council and the Grant No. 9610645 from the City University of Hong Kong. Y.H. also thanks the Mainz Institute for Theoretical Physics (MITP) of the PRISMA+ Cluster of Excellence
(Project ID 390831469) for its hospitality and partial support during the completion of this work. P.Y. acknowledeges support from the Natural
Science Foundation of Jiangsu Province (Grant No. BK20241255).

\bibliographystyle{apsrev4-1}

\bibliography{biblio.bib}

\begin{thebibliography}{68}%
\makeatletter
\providecommand \@ifxundefined [1]{%
 \@ifx{#1\undefined}
}%
\providecommand \@ifnum [1]{%
 \ifnum #1\expandafter \@firstoftwo
 \else \expandafter \@secondoftwo
 \fi
}%
\providecommand \@ifx [1]{%
 \ifx #1\expandafter \@firstoftwo
 \else \expandafter \@secondoftwo
 \fi
}%
\providecommand \natexlab [1]{#1}%
\providecommand \enquote  [1]{``#1''}%
\providecommand \bibnamefont  [1]{#1}%
\providecommand \bibfnamefont [1]{#1}%
\providecommand \citenamefont [1]{#1}%
\providecommand \href@noop [0]{\@secondoftwo}%
\providecommand \href [0]{\begingroup \@sanitize@url \@href}%
\providecommand \@href[1]{\@@startlink{#1}\@@href}%
\providecommand \@@href[1]{\endgroup#1\@@endlink}%
\providecommand \@sanitize@url [0]{\catcode `\\12\catcode `\$12\catcode
  `\&12\catcode `\#12\catcode `\^12\catcode `\_12\catcode `\%12\relax}%
\providecommand \@@startlink[1]{}%
\providecommand \@@endlink[0]{}%
\providecommand \url  [0]{\begingroup\@sanitize@url \@url }%
\providecommand \@url [1]{\endgroup\@href {#1}{\urlprefix }}%
\providecommand \urlprefix  [0]{URL }%
\providecommand \Eprint [0]{\href }%
\providecommand \doibase [0]{http://dx.doi.org/}%
\providecommand \selectlanguage [0]{\@gobble}%
\providecommand \bibinfo  [0]{\@secondoftwo}%
\providecommand \bibfield  [0]{\@secondoftwo}%
\providecommand \translation [1]{[#1]}%
\providecommand \BibitemOpen [0]{}%
\providecommand \bibitemStop [0]{}%
\providecommand \bibitemNoStop [0]{.\EOS\space}%
\providecommand \EOS [0]{\spacefactor3000\relax}%
\providecommand \BibitemShut  [1]{\csname bibitem#1\endcsname}%
\let\auto@bib@innerbib\@empty
\bibitem [{\citenamefont {Peccei}\ and\ \citenamefont
  {Quinn}(1977)}]{Peccei:1977hh}%
  \BibitemOpen
  \bibfield  {author} {\bibinfo {author} {\bibfnamefont {R.~D.}\ \bibnamefont
  {Peccei}}\ and\ \bibinfo {author} {\bibfnamefont {H.~R.}\ \bibnamefont
  {Quinn}},\ }\href {\doibase 10.1103/PhysRevLett.38.1440} {\bibfield
  {journal} {\bibinfo  {journal} {Phys. Rev. Lett.}\ }\textbf {\bibinfo
  {volume} {38}},\ \bibinfo {pages} {1440} (\bibinfo {year}
  {1977})}\BibitemShut {NoStop}%
\bibitem [{\citenamefont {Weinberg}(1978)}]{Weinberg:1977ma}%
  \BibitemOpen
  \bibfield  {author} {\bibinfo {author} {\bibfnamefont {S.}~\bibnamefont
  {Weinberg}},\ }\href {\doibase 10.1103/PhysRevLett.40.223} {\bibfield
  {journal} {\bibinfo  {journal} {Phys. Rev. Lett.}\ }\textbf {\bibinfo
  {volume} {40}},\ \bibinfo {pages} {223} (\bibinfo {year} {1978})}\BibitemShut
  {NoStop}%
\bibitem [{\citenamefont {Wilczek}(1978)}]{Wilczek:1977pj}%
  \BibitemOpen
  \bibfield  {author} {\bibinfo {author} {\bibfnamefont {F.}~\bibnamefont
  {Wilczek}},\ }\href {\doibase 10.1103/PhysRevLett.40.279} {\bibfield
  {journal} {\bibinfo  {journal} {Phys. Rev. Lett.}\ }\textbf {\bibinfo
  {volume} {40}},\ \bibinfo {pages} {279} (\bibinfo {year} {1978})}\BibitemShut
  {NoStop}%
\bibitem [{\citenamefont {Svrcek}\ and\ \citenamefont
  {Witten}(2006)}]{Svrcek_2006}%
  \BibitemOpen
  \bibfield  {author} {\bibinfo {author} {\bibfnamefont {P.}~\bibnamefont
  {Svrcek}}\ and\ \bibinfo {author} {\bibfnamefont {E.}~\bibnamefont
  {Witten}},\ }\href {\doibase 10.1088/1126-6708/2006/06/051} {\bibfield
  {journal} {\bibinfo  {journal} {Journal of High Energy Physics}\ }\textbf
  {\bibinfo {volume} {2006}},\ \bibinfo {pages} {051–051} (\bibinfo {year}
  {2006})}\BibitemShut {NoStop}%
\bibitem [{\citenamefont {Abbott}\ and\ \citenamefont
  {Sikivie}(1983)}]{Abbott:1982af}%
  \BibitemOpen
  \bibfield  {author} {\bibinfo {author} {\bibfnamefont {L.~F.}\ \bibnamefont
  {Abbott}}\ and\ \bibinfo {author} {\bibfnamefont {P.}~\bibnamefont
  {Sikivie}},\ }\href {\doibase 10.1016/0370-2693(83)90638-X} {\bibfield
  {journal} {\bibinfo  {journal} {Phys. Lett. B}\ }\textbf {\bibinfo {volume}
  {120}},\ \bibinfo {pages} {133} (\bibinfo {year} {1983})}\BibitemShut
  {NoStop}%
\bibitem [{\citenamefont {Preskill}\ \emph {et~al.}(1983)\citenamefont
  {Preskill}, \citenamefont {Wise},\ and\ \citenamefont
  {Wilczek}}]{Preskill:1982cy}%
  \BibitemOpen
  \bibfield  {author} {\bibinfo {author} {\bibfnamefont {J.}~\bibnamefont
  {Preskill}}, \bibinfo {author} {\bibfnamefont {M.~B.}\ \bibnamefont {Wise}},
  \ and\ \bibinfo {author} {\bibfnamefont {F.}~\bibnamefont {Wilczek}},\ }\href
  {\doibase 10.1016/0370-2693(83)90637-8} {\bibfield  {journal} {\bibinfo
  {journal} {Phys. Lett. B}\ }\textbf {\bibinfo {volume} {120}},\ \bibinfo
  {pages} {127} (\bibinfo {year} {1983})}\BibitemShut {NoStop}%
\bibitem [{\citenamefont {Dine}\ and\ \citenamefont
  {Fischler}(1983)}]{Dine:1982ah}%
  \BibitemOpen
  \bibfield  {author} {\bibinfo {author} {\bibfnamefont {M.}~\bibnamefont
  {Dine}}\ and\ \bibinfo {author} {\bibfnamefont {W.}~\bibnamefont
  {Fischler}},\ }\href {\doibase 10.1016/0370-2693(83)90639-1} {\bibfield
  {journal} {\bibinfo  {journal} {Phys. Lett. B}\ }\textbf {\bibinfo {volume}
  {120}},\ \bibinfo {pages} {137} (\bibinfo {year} {1983})}\BibitemShut
  {NoStop}%
\bibitem [{\citenamefont {de~Salas}\ and\ \citenamefont
  {Widmark}(2021)}]{de_Salas_2021}%
  \BibitemOpen
  \bibfield  {author} {\bibinfo {author} {\bibfnamefont {P.~F.}\ \bibnamefont
  {de~Salas}}\ and\ \bibinfo {author} {\bibfnamefont {A.}~\bibnamefont
  {Widmark}},\ }\href {\doibase 10.1088/1361-6633/ac24e7} {\bibfield  {journal}
  {\bibinfo  {journal} {Reports on Progress in Physics}\ }\textbf {\bibinfo
  {volume} {84}},\ \bibinfo {pages} {104901} (\bibinfo {year}
  {2021})}\BibitemShut {NoStop}%
\bibitem [{\citenamefont {Abel}\ \emph {et~al.}(2017)\citenamefont {Abel} \emph
  {et~al.}}]{Abel:2017rtm}%
  \BibitemOpen
  \bibfield  {author} {\bibinfo {author} {\bibfnamefont {C.}~\bibnamefont
  {Abel}} \emph {et~al.},\ }\href {\doibase 10.1103/PhysRevX.7.041034}
  {\bibfield  {journal} {\bibinfo  {journal} {Phys. Rev. X}\ }\textbf {\bibinfo
  {volume} {7}},\ \bibinfo {pages} {041034} (\bibinfo {year} {2017})},\ \Eprint
  {http://arxiv.org/abs/1708.06367} {arXiv:1708.06367 [hep-ph]} \BibitemShut
  {NoStop}%
\bibitem [{\citenamefont {Jackson~Kimball}\ \emph {et~al.}(2020)\citenamefont
  {Jackson~Kimball} \emph {et~al.}}]{JacksonKimball:2017elr}%
  \BibitemOpen
  \bibfield  {author} {\bibinfo {author} {\bibfnamefont {D.~F.}\ \bibnamefont
  {Jackson~Kimball}} \emph {et~al.},\ }\href {\doibase
  10.1007/978-3-030-43761-9_13} {\bibfield  {journal} {\bibinfo  {journal}
  {Springer Proc. Phys.}\ }\textbf {\bibinfo {volume} {245}},\ \bibinfo {pages}
  {105} (\bibinfo {year} {2020})},\ \Eprint {http://arxiv.org/abs/1711.08999}
  {arXiv:1711.08999 [physics.ins-det]} \BibitemShut {NoStop}%
\bibitem [{\citenamefont {Wu}\ \emph {et~al.}(2019)\citenamefont {Wu},
  \citenamefont {Blanchard}, \citenamefont {Centers}, \citenamefont {Figueroa},
  \citenamefont {Garcon}, \citenamefont {Graham}, \citenamefont {Kimball},
  \citenamefont {Rajendran}, \citenamefont {Stadnik}, \citenamefont {Sushkov},
  \citenamefont {Wickenbrock},\ and\ \citenamefont {Budker}}]{Wu_2019}%
  \BibitemOpen
  \bibfield  {author} {\bibinfo {author} {\bibfnamefont {T.}~\bibnamefont
  {Wu}}, \bibinfo {author} {\bibfnamefont {J.~W.}\ \bibnamefont {Blanchard}},
  \bibinfo {author} {\bibfnamefont {G.~P.}\ \bibnamefont {Centers}}, \bibinfo
  {author} {\bibfnamefont {N.~L.}\ \bibnamefont {Figueroa}}, \bibinfo {author}
  {\bibfnamefont {A.}~\bibnamefont {Garcon}}, \bibinfo {author} {\bibfnamefont
  {P.~W.}\ \bibnamefont {Graham}}, \bibinfo {author} {\bibfnamefont {D.~F.~J.}\
  \bibnamefont {Kimball}}, \bibinfo {author} {\bibfnamefont {S.}~\bibnamefont
  {Rajendran}}, \bibinfo {author} {\bibfnamefont {Y.~V.}\ \bibnamefont
  {Stadnik}}, \bibinfo {author} {\bibfnamefont {A.~O.}\ \bibnamefont
  {Sushkov}}, \bibinfo {author} {\bibfnamefont {A.}~\bibnamefont
  {Wickenbrock}}, \ and\ \bibinfo {author} {\bibfnamefont {D.}~\bibnamefont
  {Budker}},\ }\href {\doibase 10.1103/physrevlett.122.191302} {\bibfield
  {journal} {\bibinfo  {journal} {Physical Review Letters}\ }\textbf {\bibinfo
  {volume} {122}} (\bibinfo {year} {2019}),\
  10.1103/physrevlett.122.191302}\BibitemShut {NoStop}%
\bibitem [{\citenamefont {Garcon}\ \emph {et~al.}(2019)\citenamefont {Garcon}
  \emph {et~al.}}]{Garcon:2019inh}%
  \BibitemOpen
  \bibfield  {author} {\bibinfo {author} {\bibfnamefont {A.}~\bibnamefont
  {Garcon}} \emph {et~al.},\ }\href {\doibase 10.1126/sciadv.aax4539}
  {\bibfield  {journal} {\bibinfo  {journal} {Sci. Adv.}\ }\textbf {\bibinfo
  {volume} {5}},\ \bibinfo {pages} {eaax4539} (\bibinfo {year} {2019})},\
  \Eprint {http://arxiv.org/abs/1902.04644} {arXiv:1902.04644 [hep-ex]}
  \BibitemShut {NoStop}%
\bibitem [{\citenamefont {Bloch}\ \emph {et~al.}(2020)\citenamefont {Bloch},
  \citenamefont {Hochberg}, \citenamefont {Kuflik},\ and\ \citenamefont
  {Volansky}}]{Bloch:2019lcy}%
  \BibitemOpen
  \bibfield  {author} {\bibinfo {author} {\bibfnamefont {I.~M.}\ \bibnamefont
  {Bloch}}, \bibinfo {author} {\bibfnamefont {Y.}~\bibnamefont {Hochberg}},
  \bibinfo {author} {\bibfnamefont {E.}~\bibnamefont {Kuflik}}, \ and\ \bibinfo
  {author} {\bibfnamefont {T.}~\bibnamefont {Volansky}},\ }\href {\doibase
  10.1007/JHEP01(2020)167} {\bibfield  {journal} {\bibinfo  {journal} {JHEP}\
  }\textbf {\bibinfo {volume} {01}},\ \bibinfo {pages} {167} (\bibinfo {year}
  {2020})},\ \Eprint {http://arxiv.org/abs/1907.03767} {arXiv:1907.03767
  [hep-ph]} \BibitemShut {NoStop}%
\bibitem [{\citenamefont {Bloch}\ \emph {et~al.}(2022)\citenamefont {Bloch},
  \citenamefont {Ronen}, \citenamefont {Shaham}, \citenamefont {Katz},
  \citenamefont {Volansky},\ and\ \citenamefont {Katz}}]{Bloch:2021vnn}%
  \BibitemOpen
  \bibfield  {author} {\bibinfo {author} {\bibfnamefont {I.~M.}\ \bibnamefont
  {Bloch}}, \bibinfo {author} {\bibfnamefont {G.}~\bibnamefont {Ronen}},
  \bibinfo {author} {\bibfnamefont {R.}~\bibnamefont {Shaham}}, \bibinfo
  {author} {\bibfnamefont {O.}~\bibnamefont {Katz}}, \bibinfo {author}
  {\bibfnamefont {T.}~\bibnamefont {Volansky}}, \ and\ \bibinfo {author}
  {\bibfnamefont {O.}~\bibnamefont {Katz}} (\bibinfo {collaboration}
  {NASDUCK}),\ }\href {\doibase 10.1126/sciadv.abl8919} {\bibfield  {journal}
  {\bibinfo  {journal} {Sci. Adv.}\ }\textbf {\bibinfo {volume} {8}},\ \bibinfo
  {pages} {abl8919} (\bibinfo {year} {2022})},\ \Eprint
  {http://arxiv.org/abs/2105.04603} {arXiv:2105.04603 [hep-ph]} \BibitemShut
  {NoStop}%
\bibitem [{\citenamefont {Jiang}\ \emph {et~al.}(2021)\citenamefont {Jiang},
  \citenamefont {Su}, \citenamefont {Garcon}, \citenamefont {Peng},\ and\
  \citenamefont {Budker}}]{Jiang:2021dby}%
  \BibitemOpen
  \bibfield  {author} {\bibinfo {author} {\bibfnamefont {M.}~\bibnamefont
  {Jiang}}, \bibinfo {author} {\bibfnamefont {H.}~\bibnamefont {Su}}, \bibinfo
  {author} {\bibfnamefont {A.}~\bibnamefont {Garcon}}, \bibinfo {author}
  {\bibfnamefont {X.}~\bibnamefont {Peng}}, \ and\ \bibinfo {author}
  {\bibfnamefont {D.}~\bibnamefont {Budker}},\ }\href {\doibase
  10.1038/s41567-021-01392-z} {\bibfield  {journal} {\bibinfo  {journal}
  {Nature Phys.}\ }\textbf {\bibinfo {volume} {17}},\ \bibinfo {pages} {1402}
  (\bibinfo {year} {2021})},\ \Eprint {http://arxiv.org/abs/2102.01448}
  {arXiv:2102.01448 [hep-ph]} \BibitemShut {NoStop}%
\bibitem [{\citenamefont {Bloch}\ \emph {et~al.}(2023)\citenamefont {Bloch},
  \citenamefont {Shaham}, \citenamefont {Hochberg}, \citenamefont {Kuflik},
  \citenamefont {Volansky},\ and\ \citenamefont {Katz}}]{Bloch:2022kjm}%
  \BibitemOpen
  \bibfield  {author} {\bibinfo {author} {\bibfnamefont {I.~M.}\ \bibnamefont
  {Bloch}}, \bibinfo {author} {\bibfnamefont {R.}~\bibnamefont {Shaham}},
  \bibinfo {author} {\bibfnamefont {Y.}~\bibnamefont {Hochberg}}, \bibinfo
  {author} {\bibfnamefont {E.}~\bibnamefont {Kuflik}}, \bibinfo {author}
  {\bibfnamefont {T.}~\bibnamefont {Volansky}}, \ and\ \bibinfo {author}
  {\bibfnamefont {O.}~\bibnamefont {Katz}} (\bibinfo {collaboration}
  {NASDUCK}),\ }\href {\doibase 10.1038/s41467-023-41162-4} {\bibfield
  {journal} {\bibinfo  {journal} {Nature Commun.}\ }\textbf {\bibinfo {volume}
  {14}},\ \bibinfo {pages} {5784} (\bibinfo {year} {2023})},\ \Eprint
  {http://arxiv.org/abs/2209.13588} {arXiv:2209.13588 [hep-ph]} \BibitemShut
  {NoStop}%
\bibitem [{\citenamefont {Abel}\ \emph {et~al.}(2023)\citenamefont {Abel} \emph
  {et~al.}}]{Abel:2022vfg}%
  \BibitemOpen
  \bibfield  {author} {\bibinfo {author} {\bibfnamefont {C.}~\bibnamefont
  {Abel}} \emph {et~al.},\ }\href {\doibase 10.21468/SciPostPhys.15.2.058}
  {\bibfield  {journal} {\bibinfo  {journal} {SciPost Phys.}\ }\textbf
  {\bibinfo {volume} {15}},\ \bibinfo {pages} {058} (\bibinfo {year} {2023})},\
  \Eprint {http://arxiv.org/abs/2212.02403} {arXiv:2212.02403 [nucl-ex]}
  \BibitemShut {NoStop}%
\bibitem [{\citenamefont {Gao}\ \emph {et~al.}(2022)\citenamefont {Gao},
  \citenamefont {Halperin}, \citenamefont {Kahn}, \citenamefont {Nguyen},
  \citenamefont {Sch{\"u}tte-Engel},\ and\ \citenamefont
  {Scott}}]{Gao:2022nuq}%
  \BibitemOpen
  \bibfield  {author} {\bibinfo {author} {\bibfnamefont {C.}~\bibnamefont
  {Gao}}, \bibinfo {author} {\bibfnamefont {W.}~\bibnamefont {Halperin}},
  \bibinfo {author} {\bibfnamefont {Y.}~\bibnamefont {Kahn}}, \bibinfo {author}
  {\bibfnamefont {M.}~\bibnamefont {Nguyen}}, \bibinfo {author} {\bibfnamefont
  {J.}~\bibnamefont {Sch{\"u}tte-Engel}}, \ and\ \bibinfo {author}
  {\bibfnamefont {J.~W.}\ \bibnamefont {Scott}},\ }\href {\doibase
  10.1103/PhysRevLett.129.211801} {\bibfield  {journal} {\bibinfo  {journal}
  {Phys. Rev. Lett.}\ }\textbf {\bibinfo {volume} {129}},\ \bibinfo {pages}
  {211801} (\bibinfo {year} {2022})},\ \Eprint
  {http://arxiv.org/abs/2208.14454} {arXiv:2208.14454 [hep-ph]} \BibitemShut
  {NoStop}%
\bibitem [{\citenamefont {Lee}\ \emph {et~al.}(2023)\citenamefont {Lee},
  \citenamefont {Lisanti}, \citenamefont {Terrano},\ and\ \citenamefont
  {Romalis}}]{Lee:2022vvb}%
  \BibitemOpen
  \bibfield  {author} {\bibinfo {author} {\bibfnamefont {J.}~\bibnamefont
  {Lee}}, \bibinfo {author} {\bibfnamefont {M.}~\bibnamefont {Lisanti}},
  \bibinfo {author} {\bibfnamefont {W.~A.}\ \bibnamefont {Terrano}}, \ and\
  \bibinfo {author} {\bibfnamefont {M.}~\bibnamefont {Romalis}},\ }\href
  {\doibase 10.1103/PhysRevX.13.011050} {\bibfield  {journal} {\bibinfo
  {journal} {Phys. Rev. X}\ }\textbf {\bibinfo {volume} {13}},\ \bibinfo
  {pages} {011050} (\bibinfo {year} {2023})},\ \Eprint
  {http://arxiv.org/abs/2209.03289} {arXiv:2209.03289 [hep-ph]} \BibitemShut
  {NoStop}%
\bibitem [{\citenamefont {Wei}\ \emph {et~al.}(2025)\citenamefont {Wei} \emph
  {et~al.}}]{Wei:2023rzs}%
  \BibitemOpen
  \bibfield  {author} {\bibinfo {author} {\bibfnamefont {K.}~\bibnamefont
  {Wei}} \emph {et~al.},\ }\href {\doibase 10.1088/1361-6633/adca52} {\bibfield
   {journal} {\bibinfo  {journal} {Rept. Prog. Phys.}\ }\textbf {\bibinfo
  {volume} {88}},\ \bibinfo {pages} {057801} (\bibinfo {year} {2025})},\
  \Eprint {http://arxiv.org/abs/2306.08039} {arXiv:2306.08039 [hep-ph]}
  \BibitemShut {NoStop}%
\bibitem [{\citenamefont {Xu}\ \emph {et~al.}(2024)\citenamefont {Xu} \emph
  {et~al.}}]{Xu:2023vfn}%
  \BibitemOpen
  \bibfield  {author} {\bibinfo {author} {\bibfnamefont {Z.}~\bibnamefont {Xu}}
  \emph {et~al.},\ }\href {\doibase 10.1038/s42005-024-01713-7} {\bibfield
  {journal} {\bibinfo  {journal} {Commun. Phys.}\ }\textbf {\bibinfo {volume}
  {7}},\ \bibinfo {pages} {226} (\bibinfo {year} {2024})},\ \Eprint
  {http://arxiv.org/abs/2309.16600} {arXiv:2309.16600 [hep-ph]} \BibitemShut
  {NoStop}%
\bibitem [{\citenamefont {Gavilan-Martin}\ \emph {et~al.}(2025)\citenamefont
  {Gavilan-Martin} \emph {et~al.}}]{Gavilan-Martin:2024nlo}%
  \BibitemOpen
  \bibfield  {author} {\bibinfo {author} {\bibfnamefont {D.}~\bibnamefont
  {Gavilan-Martin}} \emph {et~al.},\ }\href {\doibase
  10.1038/s41467-025-60178-6} {\bibfield  {journal} {\bibinfo  {journal}
  {Nature Commun.}\ }\textbf {\bibinfo {volume} {16}},\ \bibinfo {pages} {4953}
  (\bibinfo {year} {2025})},\ \Eprint {http://arxiv.org/abs/2408.02668}
  {arXiv:2408.02668 [hep-ph]} \BibitemShut {NoStop}%
\bibitem [{\citenamefont {Rosenberg}\ \emph {et~al.}(2024)\citenamefont
  {Rosenberg}, \citenamefont {Rybka},\ and\ \citenamefont
  {Safdi}}]{Rosenberg2024Axions}%
  \BibitemOpen
  \bibfield  {author} {\bibinfo {author} {\bibfnamefont {L.~J.}\ \bibnamefont
  {Rosenberg}}, \bibinfo {author} {\bibfnamefont {G.}~\bibnamefont {Rybka}}, \
  and\ \bibinfo {author} {\bibfnamefont {B.}~\bibnamefont {Safdi}} (\bibinfo
  {collaboration} {Particle Data Group}),\ }in\ \href
  {https://pdg.lbl.gov/2024/reviews/rpp2024-rev-axions.pdf} {\emph {\bibinfo
  {booktitle} {Review of Particle Physics}}},\ Vol.\ \bibinfo {volume} {110},\
  \bibinfo {editor} {edited by\ \bibinfo {editor} {\bibfnamefont
  {S.}~\bibnamefont {Navas}} \emph {et~al.}}\ (\bibinfo {year} {2024})\ p.\
  \bibinfo {pages} {030001},\ \bibinfo {note} {revised August 2023}\BibitemShut
  {NoStop}%
\bibitem [{\citenamefont {Evans}\ \emph {et~al.}(2019)\citenamefont {Evans},
  \citenamefont {O'Hare},\ and\ \citenamefont {McCabe}}]{Evans:2018bqy}%
  \BibitemOpen
  \bibfield  {author} {\bibinfo {author} {\bibfnamefont {N.~W.}\ \bibnamefont
  {Evans}}, \bibinfo {author} {\bibfnamefont {C.~A.~J.}\ \bibnamefont
  {O'Hare}}, \ and\ \bibinfo {author} {\bibfnamefont {C.}~\bibnamefont
  {McCabe}},\ }\href {\doibase 10.1103/PhysRevD.99.023012} {\bibfield
  {journal} {\bibinfo  {journal} {Phys. Rev. D}\ }\textbf {\bibinfo {volume}
  {99}},\ \bibinfo {pages} {023012} (\bibinfo {year} {2019})},\ \Eprint
  {http://arxiv.org/abs/1810.11468} {arXiv:1810.11468 [astro-ph.GA]}
  \BibitemShut {NoStop}%
\bibitem [{\citenamefont {Capozzi}\ and\ \citenamefont
  {Raffelt}(2020)}]{Capozzi:2020cbu}%
  \BibitemOpen
  \bibfield  {author} {\bibinfo {author} {\bibfnamefont {F.}~\bibnamefont
  {Capozzi}}\ and\ \bibinfo {author} {\bibfnamefont {G.}~\bibnamefont
  {Raffelt}},\ }\href {\doibase 10.1103/PhysRevD.102.083007} {\bibfield
  {journal} {\bibinfo  {journal} {Phys. Rev. D}\ }\textbf {\bibinfo {volume}
  {102}},\ \bibinfo {pages} {083007} (\bibinfo {year} {2020})},\ \Eprint
  {http://arxiv.org/abs/2007.03694} {arXiv:2007.03694 [astro-ph.SR]}
  \BibitemShut {NoStop}%
\bibitem [{\citenamefont {Aprile}\ \emph {et~al.}(2022)\citenamefont {Aprile}
  \emph {et~al.}}]{XENON:2022ltv}%
  \BibitemOpen
  \bibfield  {author} {\bibinfo {author} {\bibfnamefont {E.}~\bibnamefont
  {Aprile}} \emph {et~al.} (\bibinfo {collaboration} {XENON}),\ }\href
  {\doibase 10.1103/PhysRevLett.129.161805} {\bibfield  {journal} {\bibinfo
  {journal} {Phys. Rev. Lett.}\ }\textbf {\bibinfo {volume} {129}},\ \bibinfo
  {pages} {161805} (\bibinfo {year} {2022})},\ \Eprint
  {http://arxiv.org/abs/2207.11330} {arXiv:2207.11330 [hep-ex]} \BibitemShut
  {NoStop}%
\bibitem [{\citenamefont {Gondolo}\ and\ \citenamefont
  {Raffelt}(2009)}]{Gondolo:2008dd}%
  \BibitemOpen
  \bibfield  {author} {\bibinfo {author} {\bibfnamefont {P.}~\bibnamefont
  {Gondolo}}\ and\ \bibinfo {author} {\bibfnamefont {G.~G.}\ \bibnamefont
  {Raffelt}},\ }\href {\doibase 10.1103/PhysRevD.79.107301} {\bibfield
  {journal} {\bibinfo  {journal} {Phys. Rev. D}\ }\textbf {\bibinfo {volume}
  {79}},\ \bibinfo {pages} {107301} (\bibinfo {year} {2009})},\ \Eprint
  {http://arxiv.org/abs/0807.2926} {arXiv:0807.2926 [astro-ph]} \BibitemShut
  {NoStop}%
\bibitem [{\citenamefont {Arisaka}\ \emph {et~al.}(2013)\citenamefont
  {Arisaka}, \citenamefont {Beltrame}, \citenamefont {Ghag}, \citenamefont
  {Kaidi}, \citenamefont {Lung}, \citenamefont {Lyashenko}, \citenamefont
  {Peccei}, \citenamefont {Smith},\ and\ \citenamefont {Ye}}]{Arisaka:2012pb}%
  \BibitemOpen
  \bibfield  {author} {\bibinfo {author} {\bibfnamefont {K.}~\bibnamefont
  {Arisaka}}, \bibinfo {author} {\bibfnamefont {P.}~\bibnamefont {Beltrame}},
  \bibinfo {author} {\bibfnamefont {C.}~\bibnamefont {Ghag}}, \bibinfo {author}
  {\bibfnamefont {J.}~\bibnamefont {Kaidi}}, \bibinfo {author} {\bibfnamefont
  {K.}~\bibnamefont {Lung}}, \bibinfo {author} {\bibfnamefont {A.}~\bibnamefont
  {Lyashenko}}, \bibinfo {author} {\bibfnamefont {R.~D.}\ \bibnamefont
  {Peccei}}, \bibinfo {author} {\bibfnamefont {P.}~\bibnamefont {Smith}}, \
  and\ \bibinfo {author} {\bibfnamefont {K.}~\bibnamefont {Ye}},\ }\href
  {\doibase 10.1016/j.astropartphys.2012.12.009} {\bibfield  {journal}
  {\bibinfo  {journal} {Astropart. Phys.}\ }\textbf {\bibinfo {volume} {44}},\
  \bibinfo {pages} {59} (\bibinfo {year} {2013})},\ \Eprint
  {http://arxiv.org/abs/1209.3810} {arXiv:1209.3810 [astro-ph.CO]} \BibitemShut
  {NoStop}%
\bibitem [{\citenamefont {Yan}\ \emph {et~al.}(2019{\natexlab{a}})\citenamefont
  {Yan}, \citenamefont {Sun}, \citenamefont {Peng}, \citenamefont {Guo},
  \citenamefont {Liu}, \citenamefont {Peng},\ and\ \citenamefont
  {Zheng}}]{Yan:2019dar}%
  \BibitemOpen
  \bibfield  {author} {\bibinfo {author} {\bibfnamefont {H.}~\bibnamefont
  {Yan}}, \bibinfo {author} {\bibfnamefont {G.~A.}\ \bibnamefont {Sun}},
  \bibinfo {author} {\bibfnamefont {S.~M.}\ \bibnamefont {Peng}}, \bibinfo
  {author} {\bibfnamefont {H.}~\bibnamefont {Guo}}, \bibinfo {author}
  {\bibfnamefont {B.~Q.}\ \bibnamefont {Liu}}, \bibinfo {author} {\bibfnamefont
  {M.}~\bibnamefont {Peng}}, \ and\ \bibinfo {author} {\bibfnamefont
  {H.}~\bibnamefont {Zheng}},\ }\href {\doibase 10.1140/epjc/s10052-019-7442-8}
  {\bibfield  {journal} {\bibinfo  {journal} {Eur. Phys. J. C}\ }\textbf
  {\bibinfo {volume} {79}},\ \bibinfo {pages} {971} (\bibinfo {year}
  {2019}{\natexlab{a}})}\BibitemShut {NoStop}%
\bibitem [{\citenamefont {Terrano}\ \emph
  {et~al.}(2015{\natexlab{a}})\citenamefont {Terrano}, \citenamefont
  {Adelberger}, \citenamefont {Lee},\ and\ \citenamefont
  {Heckel}}]{Terrano:2015sna}%
  \BibitemOpen
  \bibfield  {author} {\bibinfo {author} {\bibfnamefont {W.~A.}\ \bibnamefont
  {Terrano}}, \bibinfo {author} {\bibfnamefont {E.~G.}\ \bibnamefont
  {Adelberger}}, \bibinfo {author} {\bibfnamefont {J.~G.}\ \bibnamefont {Lee}},
  \ and\ \bibinfo {author} {\bibfnamefont {B.~R.}\ \bibnamefont {Heckel}},\
  }\href {\doibase 10.1103/PhysRevLett.115.201801} {\bibfield  {journal}
  {\bibinfo  {journal} {Phys. Rev. Lett.}\ }\textbf {\bibinfo {volume} {115}},\
  \bibinfo {pages} {201801} (\bibinfo {year} {2015}{\natexlab{a}})},\ \Eprint
  {http://arxiv.org/abs/1508.02463} {arXiv:1508.02463 [hep-ex]} \BibitemShut
  {NoStop}%
\bibitem [{\citenamefont {Terrano}\ \emph {et~al.}(2019)\citenamefont
  {Terrano}, \citenamefont {Adelberger}, \citenamefont {Hagedorn},\ and\
  \citenamefont {Heckel}}]{Terrano:2019clh}%
  \BibitemOpen
  \bibfield  {author} {\bibinfo {author} {\bibfnamefont {W.~A.}\ \bibnamefont
  {Terrano}}, \bibinfo {author} {\bibfnamefont {E.~G.}\ \bibnamefont
  {Adelberger}}, \bibinfo {author} {\bibfnamefont {C.~A.}\ \bibnamefont
  {Hagedorn}}, \ and\ \bibinfo {author} {\bibfnamefont {B.~R.}\ \bibnamefont
  {Heckel}},\ }\href {\doibase 10.1103/PhysRevLett.122.231301} {\bibfield
  {journal} {\bibinfo  {journal} {Phys. Rev. Lett.}\ }\textbf {\bibinfo
  {volume} {122}},\ \bibinfo {pages} {231301} (\bibinfo {year} {2019})},\
  \Eprint {http://arxiv.org/abs/1902.04246} {arXiv:1902.04246 [astro-ph.CO]}
  \BibitemShut {NoStop}%
\bibitem [{\citenamefont {Brandenstein}\ \emph {et~al.}(2023)\citenamefont
  {Brandenstein}, \citenamefont {Stelzl}, \citenamefont {Gutsmiedl},
  \citenamefont {Schott}, \citenamefont {Weiler},\ and\ \citenamefont
  {Fierlinger}}]{Brandenstein:2022eif}%
  \BibitemOpen
  \bibfield  {author} {\bibinfo {author} {\bibfnamefont {C.}~\bibnamefont
  {Brandenstein}}, \bibinfo {author} {\bibfnamefont {S.}~\bibnamefont
  {Stelzl}}, \bibinfo {author} {\bibfnamefont {E.}~\bibnamefont {Gutsmiedl}},
  \bibinfo {author} {\bibfnamefont {W.}~\bibnamefont {Schott}}, \bibinfo
  {author} {\bibfnamefont {A.}~\bibnamefont {Weiler}}, \ and\ \bibinfo {author}
  {\bibfnamefont {P.}~\bibnamefont {Fierlinger}},\ }\href {\doibase
  10.1051/epjconf/202328201017} {\bibfield  {journal} {\bibinfo  {journal} {EPJ
  Web Conf.}\ }\textbf {\bibinfo {volume} {282}},\ \bibinfo {pages} {01017}
  (\bibinfo {year} {2023})},\ \Eprint {http://arxiv.org/abs/2211.08439}
  {arXiv:2211.08439 [hep-ex]} \BibitemShut {NoStop}%
\bibitem [{\citenamefont {Alonso}\ \emph {et~al.}(2019)\citenamefont {Alonso},
  \citenamefont {Blas},\ and\ \citenamefont {Wolf}}]{Alonso:2018dxy}%
  \BibitemOpen
  \bibfield  {author} {\bibinfo {author} {\bibfnamefont {R.}~\bibnamefont
  {Alonso}}, \bibinfo {author} {\bibfnamefont {D.}~\bibnamefont {Blas}}, \ and\
  \bibinfo {author} {\bibfnamefont {P.}~\bibnamefont {Wolf}},\ }\href {\doibase
  10.1007/JHEP07(2019)069} {\bibfield  {journal} {\bibinfo  {journal} {JHEP}\
  }\textbf {\bibinfo {volume} {07}},\ \bibinfo {pages} {069} (\bibinfo {year}
  {2019})},\ \Eprint {http://arxiv.org/abs/1810.00889} {arXiv:1810.00889
  [hep-ph]} \BibitemShut {NoStop}%
\bibitem [{\citenamefont {Crescini}\ \emph {et~al.}(2018)\citenamefont
  {Crescini} \emph {et~al.}}]{Crescini:2018qrz}%
  \BibitemOpen
  \bibfield  {author} {\bibinfo {author} {\bibfnamefont {N.}~\bibnamefont
  {Crescini}} \emph {et~al.},\ }\href {\doibase 10.1140/epjc/s10052-018-6163-8}
  {\bibfield  {journal} {\bibinfo  {journal} {Eur. Phys. J. C}\ }\textbf
  {\bibinfo {volume} {78}},\ \bibinfo {pages} {703} (\bibinfo {year} {2018})},\
  \bibinfo {note} {[Erratum: Eur.Phys.J.C 78, 813 (2018)]},\ \Eprint
  {http://arxiv.org/abs/1806.00310} {arXiv:1806.00310 [hep-ex]} \BibitemShut
  {NoStop}%
\bibitem [{\citenamefont {Crescini}\ \emph {et~al.}(2020)\citenamefont
  {Crescini} \emph {et~al.}}]{QUAX:2020adt}%
  \BibitemOpen
  \bibfield  {author} {\bibinfo {author} {\bibfnamefont {N.}~\bibnamefont
  {Crescini}} \emph {et~al.} (\bibinfo {collaboration} {QUAX}),\ }\href
  {\doibase 10.1103/PhysRevLett.124.171801} {\bibfield  {journal} {\bibinfo
  {journal} {Phys. Rev. Lett.}\ }\textbf {\bibinfo {volume} {124}},\ \bibinfo
  {pages} {171801} (\bibinfo {year} {2020})},\ \Eprint
  {http://arxiv.org/abs/2001.08940} {arXiv:2001.08940 [hep-ex]} \BibitemShut
  {NoStop}%
\bibitem [{\citenamefont {Terrano}\ \emph
  {et~al.}(2015{\natexlab{b}})\citenamefont {Terrano}, \citenamefont
  {Adelberger}, \citenamefont {Lee},\ and\ \citenamefont
  {Heckel}}]{terrano2015shortrangespindependentinteractionselectrons}%
  \BibitemOpen
  \bibfield  {author} {\bibinfo {author} {\bibfnamefont {W.~A.}\ \bibnamefont
  {Terrano}}, \bibinfo {author} {\bibfnamefont {E.~G.}\ \bibnamefont
  {Adelberger}}, \bibinfo {author} {\bibfnamefont {J.~G.}\ \bibnamefont {Lee}},
  \ and\ \bibinfo {author} {\bibfnamefont {B.~R.}\ \bibnamefont {Heckel}},\
  }\href {https://arxiv.org/abs/1508.02463} {\enquote {\bibinfo {title}
  {Short-range spin-dependent interactions of electrons: a probe for exotic
  pseudo-goldstone bosons},}\ } (\bibinfo {year} {2015}{\natexlab{b}}),\
  \Eprint {http://arxiv.org/abs/1508.02463} {arXiv:1508.02463 [hep-ex]}
  \BibitemShut {NoStop}%
\bibitem [{\citenamefont {Ikeda}\ \emph
  {et~al.}(2022{\natexlab{a}})\citenamefont {Ikeda}, \citenamefont {Ito},
  \citenamefont {Miuchi}, \citenamefont {Soda}, \citenamefont {Kurashige},\
  and\ \citenamefont {Shikano}}]{Ikeda_2022}%
  \BibitemOpen
  \bibfield  {author} {\bibinfo {author} {\bibfnamefont {T.}~\bibnamefont
  {Ikeda}}, \bibinfo {author} {\bibfnamefont {A.}~\bibnamefont {Ito}}, \bibinfo
  {author} {\bibfnamefont {K.}~\bibnamefont {Miuchi}}, \bibinfo {author}
  {\bibfnamefont {J.}~\bibnamefont {Soda}}, \bibinfo {author} {\bibfnamefont
  {H.}~\bibnamefont {Kurashige}}, \ and\ \bibinfo {author} {\bibfnamefont
  {Y.}~\bibnamefont {Shikano}},\ }\href {\doibase 10.1103/physrevd.105.102004}
  {\bibfield  {journal} {\bibinfo  {journal} {Physical Review D}\ }\textbf
  {\bibinfo {volume} {105}} (\bibinfo {year} {2022}{\natexlab{a}}),\
  10.1103/physrevd.105.102004}\BibitemShut {NoStop}%
\bibitem [{\citenamefont {Yan}\ \emph {et~al.}(2019{\natexlab{b}})\citenamefont
  {Yan}, \citenamefont {Sun}, \citenamefont {Peng}, \citenamefont {Guo},
  \citenamefont {Liu}, \citenamefont {Peng},\ and\ \citenamefont
  {Zheng}}]{Yan_2019}%
  \BibitemOpen
  \bibfield  {author} {\bibinfo {author} {\bibfnamefont {H.}~\bibnamefont
  {Yan}}, \bibinfo {author} {\bibfnamefont {G.~A.}\ \bibnamefont {Sun}},
  \bibinfo {author} {\bibfnamefont {S.~M.}\ \bibnamefont {Peng}}, \bibinfo
  {author} {\bibfnamefont {H.}~\bibnamefont {Guo}}, \bibinfo {author}
  {\bibfnamefont {B.~Q.}\ \bibnamefont {Liu}}, \bibinfo {author} {\bibfnamefont
  {M.}~\bibnamefont {Peng}}, \ and\ \bibinfo {author} {\bibfnamefont
  {H.}~\bibnamefont {Zheng}},\ }\href@noop {} {\bibfield  {journal} {\bibinfo
  {journal} {The European Physical Journal C}\ }\textbf {\bibinfo {volume}
  {79}},\ \bibinfo {pages} {971} (\bibinfo {year}
  {2019}{\natexlab{b}})}\BibitemShut {NoStop}%
\bibitem [{\citenamefont {Flower}\ \emph {et~al.}(2019)\citenamefont {Flower},
  \citenamefont {Bourhill}, \citenamefont {Goryachev},\ and\ \citenamefont
  {Tobar}}]{Flower:2018qgb}%
  \BibitemOpen
  \bibfield  {author} {\bibinfo {author} {\bibfnamefont {G.}~\bibnamefont
  {Flower}}, \bibinfo {author} {\bibfnamefont {J.}~\bibnamefont {Bourhill}},
  \bibinfo {author} {\bibfnamefont {M.}~\bibnamefont {Goryachev}}, \ and\
  \bibinfo {author} {\bibfnamefont {M.~E.}\ \bibnamefont {Tobar}},\ }\href
  {\doibase 10.1016/j.dark.2019.100306} {\bibfield  {journal} {\bibinfo
  {journal} {Phys. Dark Univ.}\ }\textbf {\bibinfo {volume} {25}},\ \bibinfo
  {pages} {100306} (\bibinfo {year} {2019})},\ \Eprint
  {http://arxiv.org/abs/1811.09348} {arXiv:1811.09348 [physics.ins-det]}
  \BibitemShut {NoStop}%
\bibitem [{\citenamefont {Chigusa}\ \emph {et~al.}(2020)\citenamefont
  {Chigusa}, \citenamefont {Moroi},\ and\ \citenamefont
  {Nakayama}}]{Chigusa:2020gfs}%
  \BibitemOpen
  \bibfield  {author} {\bibinfo {author} {\bibfnamefont {S.}~\bibnamefont
  {Chigusa}}, \bibinfo {author} {\bibfnamefont {T.}~\bibnamefont {Moroi}}, \
  and\ \bibinfo {author} {\bibfnamefont {K.}~\bibnamefont {Nakayama}},\ }\href
  {\doibase 10.1103/PhysRevD.101.096013} {\bibfield  {journal} {\bibinfo
  {journal} {Phys. Rev. D}\ }\textbf {\bibinfo {volume} {101}},\ \bibinfo
  {pages} {096013} (\bibinfo {year} {2020})},\ \Eprint
  {http://arxiv.org/abs/2001.10666} {arXiv:2001.10666 [hep-ph]} \BibitemShut
  {NoStop}%
\bibitem [{\citenamefont {Ikeda}\ \emph
  {et~al.}(2022{\natexlab{b}})\citenamefont {Ikeda}, \citenamefont {Ito},
  \citenamefont {Miuchi}, \citenamefont {Soda}, \citenamefont {Kurashige},\
  and\ \citenamefont {Shikano}}]{Ikeda:2021mlv}%
  \BibitemOpen
  \bibfield  {author} {\bibinfo {author} {\bibfnamefont {T.}~\bibnamefont
  {Ikeda}}, \bibinfo {author} {\bibfnamefont {A.}~\bibnamefont {Ito}}, \bibinfo
  {author} {\bibfnamefont {K.}~\bibnamefont {Miuchi}}, \bibinfo {author}
  {\bibfnamefont {J.}~\bibnamefont {Soda}}, \bibinfo {author} {\bibfnamefont
  {H.}~\bibnamefont {Kurashige}}, \ and\ \bibinfo {author} {\bibfnamefont
  {Y.}~\bibnamefont {Shikano}},\ }\href {\doibase 10.1103/PhysRevD.105.102004}
  {\bibfield  {journal} {\bibinfo  {journal} {Phys. Rev. D}\ }\textbf {\bibinfo
  {volume} {105}},\ \bibinfo {pages} {102004} (\bibinfo {year}
  {2022}{\natexlab{b}})},\ \Eprint {http://arxiv.org/abs/2102.08764}
  {arXiv:2102.08764 [hep-ex]} \BibitemShut {NoStop}%
\bibitem [{\citenamefont {Chu}\ \emph {et~al.}(2019)\citenamefont {Chu},
  \citenamefont {Kim},\ and\ \citenamefont {Savukov}}]{Chu:2019iry}%
  \BibitemOpen
  \bibfield  {author} {\bibinfo {author} {\bibfnamefont {P.~H.}\ \bibnamefont
  {Chu}}, \bibinfo {author} {\bibfnamefont {Y.~J.}\ \bibnamefont {Kim}}, \ and\
  \bibinfo {author} {\bibfnamefont {I.}~\bibnamefont {Savukov}},\ }\href@noop
  {} {\  (\bibinfo {year} {2019})},\ \Eprint {http://arxiv.org/abs/1904.10543}
  {arXiv:1904.10543 [hep-ph]} \BibitemShut {NoStop}%
\bibitem [{\citenamefont {Sikivie}(2014)}]{Sikivie:2014lha}%
  \BibitemOpen
  \bibfield  {author} {\bibinfo {author} {\bibfnamefont {P.}~\bibnamefont
  {Sikivie}},\ }\href {\doibase 10.1103/PhysRevLett.113.201301} {\bibfield
  {journal} {\bibinfo  {journal} {Phys. Rev. Lett.}\ }\textbf {\bibinfo
  {volume} {113}},\ \bibinfo {pages} {201301} (\bibinfo {year} {2014})},\
  \bibinfo {note} {[Erratum: Phys.Rev.Lett. 125, 029901 (2020)]},\ \Eprint
  {http://arxiv.org/abs/1409.2806} {arXiv:1409.2806 [hep-ph]} \BibitemShut
  {NoStop}%
\bibitem [{\citenamefont {Hochberg}\ \emph {et~al.}(2017)\citenamefont
  {Hochberg}, \citenamefont {Lin},\ and\ \citenamefont
  {Zurek}}]{Hochberg:2016sqx}%
  \BibitemOpen
  \bibfield  {author} {\bibinfo {author} {\bibfnamefont {Y.}~\bibnamefont
  {Hochberg}}, \bibinfo {author} {\bibfnamefont {T.}~\bibnamefont {Lin}}, \
  and\ \bibinfo {author} {\bibfnamefont {K.~M.}\ \bibnamefont {Zurek}},\ }\href
  {\doibase 10.1103/PhysRevD.95.023013} {\bibfield  {journal} {\bibinfo
  {journal} {Phys. Rev. D}\ }\textbf {\bibinfo {volume} {95}},\ \bibinfo
  {pages} {023013} (\bibinfo {year} {2017})},\ \Eprint
  {http://arxiv.org/abs/1608.01994} {arXiv:1608.01994 [hep-ph]} \BibitemShut
  {NoStop}%
\bibitem [{\citenamefont {Hochberg}\ \emph {et~al.}(2016)\citenamefont
  {Hochberg}, \citenamefont {Lin},\ and\ \citenamefont
  {Zurek}}]{Hochberg:2016ajh}%
  \BibitemOpen
  \bibfield  {author} {\bibinfo {author} {\bibfnamefont {Y.}~\bibnamefont
  {Hochberg}}, \bibinfo {author} {\bibfnamefont {T.}~\bibnamefont {Lin}}, \
  and\ \bibinfo {author} {\bibfnamefont {K.~M.}\ \bibnamefont {Zurek}},\ }\href
  {\doibase 10.1103/PhysRevD.94.015019} {\bibfield  {journal} {\bibinfo
  {journal} {Phys. Rev. D}\ }\textbf {\bibinfo {volume} {94}},\ \bibinfo
  {pages} {015019} (\bibinfo {year} {2016})},\ \Eprint
  {http://arxiv.org/abs/1604.06800} {arXiv:1604.06800 [hep-ph]} \BibitemShut
  {NoStop}%
\bibitem [{\citenamefont {Arvanitaki}\ \emph {et~al.}(2018)\citenamefont
  {Arvanitaki}, \citenamefont {Dimopoulos},\ and\ \citenamefont
  {Van~Tilburg}}]{Arvanitaki:2017nhi}%
  \BibitemOpen
  \bibfield  {author} {\bibinfo {author} {\bibfnamefont {A.}~\bibnamefont
  {Arvanitaki}}, \bibinfo {author} {\bibfnamefont {S.}~\bibnamefont
  {Dimopoulos}}, \ and\ \bibinfo {author} {\bibfnamefont {K.}~\bibnamefont
  {Van~Tilburg}},\ }\href {\doibase 10.1103/PhysRevX.8.041001} {\bibfield
  {journal} {\bibinfo  {journal} {Phys. Rev. X}\ }\textbf {\bibinfo {volume}
  {8}},\ \bibinfo {pages} {041001} (\bibinfo {year} {2018})},\ \Eprint
  {http://arxiv.org/abs/1709.05354} {arXiv:1709.05354 [hep-ph]} \BibitemShut
  {NoStop}%
\bibitem [{\citenamefont {Arvanitaki}\ \emph {et~al.}(2024)\citenamefont
  {Arvanitaki}, \citenamefont {Madden},\ and\ \citenamefont
  {Van~Tilburg}}]{Arvanitaki:2021wjk}%
  \BibitemOpen
  \bibfield  {author} {\bibinfo {author} {\bibfnamefont {A.}~\bibnamefont
  {Arvanitaki}}, \bibinfo {author} {\bibfnamefont {A.}~\bibnamefont {Madden}},
  \ and\ \bibinfo {author} {\bibfnamefont {K.}~\bibnamefont {Van~Tilburg}},\
  }\href {\doibase 10.1103/PhysRevD.109.072009} {\bibfield  {journal} {\bibinfo
   {journal} {Phys. Rev. D}\ }\textbf {\bibinfo {volume} {109}},\ \bibinfo
  {pages} {072009} (\bibinfo {year} {2024})},\ \Eprint
  {http://arxiv.org/abs/2112.11466} {arXiv:2112.11466 [hep-ph]} \BibitemShut
  {NoStop}%
\bibitem [{\citenamefont {Chen}\ \emph {et~al.}(2022)\citenamefont {Chen},
  \citenamefont {Mitridate}, \citenamefont {Trickle}, \citenamefont {Zhang},
  \citenamefont {Bernardi},\ and\ \citenamefont {Zurek}}]{Chen:2022pyd}%
  \BibitemOpen
  \bibfield  {author} {\bibinfo {author} {\bibfnamefont {H.-Y.}\ \bibnamefont
  {Chen}}, \bibinfo {author} {\bibfnamefont {A.}~\bibnamefont {Mitridate}},
  \bibinfo {author} {\bibfnamefont {T.}~\bibnamefont {Trickle}}, \bibinfo
  {author} {\bibfnamefont {Z.}~\bibnamefont {Zhang}}, \bibinfo {author}
  {\bibfnamefont {M.}~\bibnamefont {Bernardi}}, \ and\ \bibinfo {author}
  {\bibfnamefont {K.~M.}\ \bibnamefont {Zurek}},\ }\href {\doibase
  10.1103/PhysRevD.106.015024} {\bibfield  {journal} {\bibinfo  {journal}
  {Phys. Rev. D}\ }\textbf {\bibinfo {volume} {106}},\ \bibinfo {pages}
  {015024} (\bibinfo {year} {2022})},\ \Eprint
  {http://arxiv.org/abs/2202.11716} {arXiv:2202.11716 [hep-ph]} \BibitemShut
  {NoStop}%
\bibitem [{\citenamefont {Chigusa}\ \emph {et~al.}(2023)\citenamefont
  {Chigusa}, \citenamefont {Moroi}, \citenamefont {Nakayama},\ and\
  \citenamefont {Sichanugrist}}]{Chigusa:2023hmz}%
  \BibitemOpen
  \bibfield  {author} {\bibinfo {author} {\bibfnamefont {S.}~\bibnamefont
  {Chigusa}}, \bibinfo {author} {\bibfnamefont {T.}~\bibnamefont {Moroi}},
  \bibinfo {author} {\bibfnamefont {K.}~\bibnamefont {Nakayama}}, \ and\
  \bibinfo {author} {\bibfnamefont {T.}~\bibnamefont {Sichanugrist}},\ }\href
  {\doibase 10.1103/PhysRevD.108.095007} {\bibfield  {journal} {\bibinfo
  {journal} {Phys. Rev. D}\ }\textbf {\bibinfo {volume} {108}},\ \bibinfo
  {pages} {095007} (\bibinfo {year} {2023})},\ \Eprint
  {http://arxiv.org/abs/2307.08577} {arXiv:2307.08577 [hep-ph]} \BibitemShut
  {NoStop}%
\bibitem [{\citenamefont {Mitridate}\ \emph {et~al.}(2024)\citenamefont
  {Mitridate}, \citenamefont {Pardo}, \citenamefont {Trickle},\ and\
  \citenamefont {Zurek}}]{Mitridate:2023izi}%
  \BibitemOpen
  \bibfield  {author} {\bibinfo {author} {\bibfnamefont {A.}~\bibnamefont
  {Mitridate}}, \bibinfo {author} {\bibfnamefont {K.}~\bibnamefont {Pardo}},
  \bibinfo {author} {\bibfnamefont {T.}~\bibnamefont {Trickle}}, \ and\
  \bibinfo {author} {\bibfnamefont {K.~M.}\ \bibnamefont {Zurek}},\ }\href
  {\doibase 10.1103/PhysRevD.109.015010} {\bibfield  {journal} {\bibinfo
  {journal} {Phys. Rev. D}\ }\textbf {\bibinfo {volume} {109}},\ \bibinfo
  {pages} {015010} (\bibinfo {year} {2024})},\ \Eprint
  {http://arxiv.org/abs/2308.06314} {arXiv:2308.06314 [hep-ph]} \BibitemShut
  {NoStop}%
\bibitem [{\citenamefont {Berlin}\ \emph {et~al.}(2024)\citenamefont {Berlin},
  \citenamefont {Millar}, \citenamefont {Trickle},\ and\ \citenamefont
  {Zhou}}]{Berlin:2023ubt}%
  \BibitemOpen
  \bibfield  {author} {\bibinfo {author} {\bibfnamefont {A.}~\bibnamefont
  {Berlin}}, \bibinfo {author} {\bibfnamefont {A.~J.}\ \bibnamefont {Millar}},
  \bibinfo {author} {\bibfnamefont {T.}~\bibnamefont {Trickle}}, \ and\
  \bibinfo {author} {\bibfnamefont {K.}~\bibnamefont {Zhou}},\ }\href {\doibase
  10.1007/JHEP05(2024)314} {\bibfield  {journal} {\bibinfo  {journal} {JHEP}\
  }\textbf {\bibinfo {volume} {05}},\ \bibinfo {pages} {314} (\bibinfo {year}
  {2024})},\ \Eprint {http://arxiv.org/abs/2312.11601} {arXiv:2312.11601
  [hep-ph]} \BibitemShut {NoStop}%
\bibitem [{\citenamefont {Chigusa}\ \emph {et~al.}(2025)\citenamefont
  {Chigusa}, \citenamefont {Hazumi}, \citenamefont {Herbschleb}, \citenamefont
  {Mizuochi},\ and\ \citenamefont {Nakayama}}]{Chigusa:2023roq}%
  \BibitemOpen
  \bibfield  {author} {\bibinfo {author} {\bibfnamefont {S.}~\bibnamefont
  {Chigusa}}, \bibinfo {author} {\bibfnamefont {M.}~\bibnamefont {Hazumi}},
  \bibinfo {author} {\bibfnamefont {E.~D.}\ \bibnamefont {Herbschleb}},
  \bibinfo {author} {\bibfnamefont {N.}~\bibnamefont {Mizuochi}}, \ and\
  \bibinfo {author} {\bibfnamefont {K.}~\bibnamefont {Nakayama}},\ }\href
  {\doibase 10.1007/JHEP03(2025)083} {\bibfield  {journal} {\bibinfo  {journal}
  {JHEP}\ }\textbf {\bibinfo {volume} {03}},\ \bibinfo {pages} {083} (\bibinfo
  {year} {2025})},\ \Eprint {http://arxiv.org/abs/2302.12756} {arXiv:2302.12756
  [hep-ph]} \BibitemShut {NoStop}%
\bibitem [{\citenamefont {Kilian}\ \emph {et~al.}(2024)\citenamefont {Kilian}
  \emph {et~al.}}]{Kilian:2024fsg}%
  \BibitemOpen
  \bibfield  {author} {\bibinfo {author} {\bibfnamefont {E.}~\bibnamefont
  {Kilian}} \emph {et~al.},\ }\href {\doibase 10.1116/5.0200916} {\bibfield
  {journal} {\bibinfo  {journal} {AVS Quantum Sci.}\ }\textbf {\bibinfo
  {volume} {6}},\ \bibinfo {pages} {030503} (\bibinfo {year} {2024})},\ \Eprint
  {http://arxiv.org/abs/2401.17990} {arXiv:2401.17990 [quant-ph]} \BibitemShut
  {NoStop}%
\bibitem [{\citenamefont {Ahrens}\ \emph {et~al.}(2025)\citenamefont {Ahrens},
  \citenamefont {Ji}, \citenamefont {Budker}, \citenamefont {Timberlake},
  \citenamefont {Ulbricht},\ and\ \citenamefont {Vinante}}]{Ahrens2025}%
  \BibitemOpen
  \bibfield  {author} {\bibinfo {author} {\bibfnamefont {F.}~\bibnamefont
  {Ahrens}}, \bibinfo {author} {\bibfnamefont {W.}~\bibnamefont {Ji}}, \bibinfo
  {author} {\bibfnamefont {D.}~\bibnamefont {Budker}}, \bibinfo {author}
  {\bibfnamefont {C.}~\bibnamefont {Timberlake}}, \bibinfo {author}
  {\bibfnamefont {H.}~\bibnamefont {Ulbricht}}, \ and\ \bibinfo {author}
  {\bibfnamefont {A.}~\bibnamefont {Vinante}},\ }\href {\doibase
  10.1103/PhysRevLett.134.110801} {\bibfield  {journal} {\bibinfo  {journal}
  {Phys. Rev. Lett.}\ }\textbf {\bibinfo {volume} {134}},\ \bibinfo {pages}
  {110801} (\bibinfo {year} {2025})}\BibitemShut {NoStop}%
\bibitem [{\citenamefont {Higgins}\ \emph {et~al.}(2024)\citenamefont
  {Higgins}, \citenamefont {Kalia},\ and\ \citenamefont
  {Liu}}]{Higgins:2023gwq}%
  \BibitemOpen
  \bibfield  {author} {\bibinfo {author} {\bibfnamefont {G.}~\bibnamefont
  {Higgins}}, \bibinfo {author} {\bibfnamefont {S.}~\bibnamefont {Kalia}}, \
  and\ \bibinfo {author} {\bibfnamefont {Z.}~\bibnamefont {Liu}},\ }\href
  {\doibase 10.1103/PhysRevD.109.055024} {\bibfield  {journal} {\bibinfo
  {journal} {Phys. Rev. D}\ }\textbf {\bibinfo {volume} {109}},\ \bibinfo
  {pages} {055024} (\bibinfo {year} {2024})},\ \Eprint
  {http://arxiv.org/abs/2310.18398} {arXiv:2310.18398 [hep-ph]} \BibitemShut
  {NoStop}%
\bibitem [{\citenamefont {Kalia}\ \emph {et~al.}(2024)\citenamefont {Kalia},
  \citenamefont {Budker}, \citenamefont {Kimball}, \citenamefont {Ji},
  \citenamefont {Liu}, \citenamefont {Sushkov}, \citenamefont {Timberlake},
  \citenamefont {Ulbricht}, \citenamefont {Vinante},\ and\ \citenamefont
  {Wang}}]{Kalia:2024eml}%
  \BibitemOpen
  \bibfield  {author} {\bibinfo {author} {\bibfnamefont {S.}~\bibnamefont
  {Kalia}}, \bibinfo {author} {\bibfnamefont {D.}~\bibnamefont {Budker}},
  \bibinfo {author} {\bibfnamefont {D.~F.~J.}\ \bibnamefont {Kimball}},
  \bibinfo {author} {\bibfnamefont {W.}~\bibnamefont {Ji}}, \bibinfo {author}
  {\bibfnamefont {Z.}~\bibnamefont {Liu}}, \bibinfo {author} {\bibfnamefont
  {A.~O.}\ \bibnamefont {Sushkov}}, \bibinfo {author} {\bibfnamefont
  {C.}~\bibnamefont {Timberlake}}, \bibinfo {author} {\bibfnamefont
  {H.}~\bibnamefont {Ulbricht}}, \bibinfo {author} {\bibfnamefont
  {A.}~\bibnamefont {Vinante}}, \ and\ \bibinfo {author} {\bibfnamefont
  {T.}~\bibnamefont {Wang}},\ }\href {\doibase 10.1103/PhysRevD.110.115029}
  {\bibfield  {journal} {\bibinfo  {journal} {Phys. Rev. D}\ }\textbf {\bibinfo
  {volume} {110}},\ \bibinfo {pages} {115029} (\bibinfo {year} {2024})},\
  \Eprint {http://arxiv.org/abs/2408.15330} {arXiv:2408.15330 [hep-ph]}
  \BibitemShut {NoStop}%
\bibitem [{\citenamefont {Jackson~Kimball}\ \emph
  {et~al.}(2016{\natexlab{a}})\citenamefont {Jackson~Kimball}, \citenamefont
  {Dudley}, \citenamefont {Li}, \citenamefont {Thulasi}, \citenamefont
  {Pustelny}, \citenamefont {Budker},\ and\ \citenamefont
  {Zolotorev}}]{JacksonKimball:2016wzv}%
  \BibitemOpen
  \bibfield  {author} {\bibinfo {author} {\bibfnamefont {D.~F.}\ \bibnamefont
  {Jackson~Kimball}}, \bibinfo {author} {\bibfnamefont {J.}~\bibnamefont
  {Dudley}}, \bibinfo {author} {\bibfnamefont {Y.}~\bibnamefont {Li}}, \bibinfo
  {author} {\bibfnamefont {S.}~\bibnamefont {Thulasi}}, \bibinfo {author}
  {\bibfnamefont {S.}~\bibnamefont {Pustelny}}, \bibinfo {author}
  {\bibfnamefont {D.}~\bibnamefont {Budker}}, \ and\ \bibinfo {author}
  {\bibfnamefont {M.}~\bibnamefont {Zolotorev}},\ }\href {\doibase
  10.1103/PhysRevD.94.082005} {\bibfield  {journal} {\bibinfo  {journal} {Phys.
  Rev. D}\ }\textbf {\bibinfo {volume} {94}},\ \bibinfo {pages} {082005}
  (\bibinfo {year} {2016}{\natexlab{a}})},\ \Eprint
  {http://arxiv.org/abs/1606.00696} {arXiv:1606.00696 [physics.ins-det]}
  \BibitemShut {NoStop}%
\bibitem [{\citenamefont {Jackson~Kimball}\ \emph
  {et~al.}(2016{\natexlab{b}})\citenamefont {Jackson~Kimball}, \citenamefont
  {Sushkov},\ and\ \citenamefont {Budker}}]{JacksonKimball2016}%
  \BibitemOpen
  \bibfield  {author} {\bibinfo {author} {\bibfnamefont {D.~F.}\ \bibnamefont
  {Jackson~Kimball}}, \bibinfo {author} {\bibfnamefont {A.~O.}\ \bibnamefont
  {Sushkov}}, \ and\ \bibinfo {author} {\bibfnamefont {D.}~\bibnamefont
  {Budker}},\ }\href {\doibase 10.1103/physrevlett.116.190801} {\bibfield
  {journal} {\bibinfo  {journal} {Phys. Rev. Lett.}\ }\textbf {\bibinfo
  {volume} {116}},\ \bibinfo {pages} {190801} (\bibinfo {year}
  {2016}{\natexlab{b}})}\BibitemShut {NoStop}%
\bibitem [{\citenamefont {Vinante}\ \emph {et~al.}(2021)\citenamefont
  {Vinante}, \citenamefont {Timberlake}, \citenamefont {Budker}, \citenamefont
  {Kimball}, \citenamefont {Sushkov},\ and\ \citenamefont
  {Ulbricht}}]{Vinante2021}%
  \BibitemOpen
  \bibfield  {author} {\bibinfo {author} {\bibfnamefont {A.}~\bibnamefont
  {Vinante}}, \bibinfo {author} {\bibfnamefont {C.}~\bibnamefont {Timberlake}},
  \bibinfo {author} {\bibfnamefont {D.}~\bibnamefont {Budker}}, \bibinfo
  {author} {\bibfnamefont {D.~F.~J.}\ \bibnamefont {Kimball}}, \bibinfo
  {author} {\bibfnamefont {A.~O.}\ \bibnamefont {Sushkov}}, \ and\ \bibinfo
  {author} {\bibfnamefont {H.}~\bibnamefont {Ulbricht}},\ }\href {\doibase
  10.1103/physrevlett.127.070801} {\bibfield  {journal} {\bibinfo  {journal}
  {Phys. Rev. Lett.}\ }\textbf {\bibinfo {volume} {127}},\ \bibinfo {pages}
  {070801} (\bibinfo {year} {2021})}\BibitemShut {NoStop}%
\bibitem [{\citenamefont {Ji}\ \emph {et~al.}(2025)\citenamefont {Ji},
  \citenamefont {Xu}, \citenamefont {Qu},\ and\ \citenamefont
  {Budker}}]{Ji:2025yvn}%
  \BibitemOpen
  \bibfield  {author} {\bibinfo {author} {\bibfnamefont {W.}~\bibnamefont
  {Ji}}, \bibinfo {author} {\bibfnamefont {C.}~\bibnamefont {Xu}}, \bibinfo
  {author} {\bibfnamefont {G.}~\bibnamefont {Qu}}, \ and\ \bibinfo {author}
  {\bibfnamefont {D.}~\bibnamefont {Budker}},\ }\href@noop {} {\  (\bibinfo
  {year} {2025})},\ \Eprint {http://arxiv.org/abs/2504.21524} {arXiv:2504.21524
  [physics.ins-det]} \BibitemShut {NoStop}%
\bibitem [{\citenamefont {Blundell}(2001)}]{blundell2001magnetism}%
  \BibitemOpen
  \bibfield  {author} {\bibinfo {author} {\bibfnamefont {S.}~\bibnamefont
  {Blundell}},\ }\href@noop {} {\emph {\bibinfo {title} {Magnetism in Condensed
  Matter}}},\ Oxford Master Series in Physics\ (\bibinfo  {publisher} {Oxford
  University Press},\ \bibinfo {address} {Oxford},\ \bibinfo {year}
  {2001})\BibitemShut {NoStop}%
\bibitem [{\citenamefont {Yin}\ \emph {et~al.}(2022)\citenamefont {Yin},
  \citenamefont {Li}, \citenamefont {Yin}, \citenamefont {Xu}, \citenamefont
  {Bian}, \citenamefont {Xie}, \citenamefont {Duan}, \citenamefont {Huang},
  \citenamefont {He},\ and\ \citenamefont {Du}}]{Yin:2022geb}%
  \BibitemOpen
  \bibfield  {author} {\bibinfo {author} {\bibfnamefont {P.}~\bibnamefont
  {Yin}}, \bibinfo {author} {\bibfnamefont {R.}~\bibnamefont {Li}}, \bibinfo
  {author} {\bibfnamefont {C.}~\bibnamefont {Yin}}, \bibinfo {author}
  {\bibfnamefont {X.}~\bibnamefont {Xu}}, \bibinfo {author} {\bibfnamefont
  {X.}~\bibnamefont {Bian}}, \bibinfo {author} {\bibfnamefont {H.}~\bibnamefont
  {Xie}}, \bibinfo {author} {\bibfnamefont {C.-K.}\ \bibnamefont {Duan}},
  \bibinfo {author} {\bibfnamefont {P.}~\bibnamefont {Huang}}, \bibinfo
  {author} {\bibfnamefont {J.-h.}\ \bibnamefont {He}}, \ and\ \bibinfo {author}
  {\bibfnamefont {J.}~\bibnamefont {Du}},\ }\href {\doibase
  10.1038/s41567-022-01706-9} {\bibfield  {journal} {\bibinfo  {journal}
  {Nature Phys.}\ }\textbf {\bibinfo {volume} {18}},\ \bibinfo {pages} {1181}
  (\bibinfo {year} {2022})}\BibitemShut {NoStop}%
\bibitem [{\citenamefont {Yin}\ \emph {et~al.}(2025)\citenamefont {Yin},
  \citenamefont {Xu}, \citenamefont {Tian}, \citenamefont {Lin}, \citenamefont
  {Sheng}, \citenamefont {Yin}, \citenamefont {Long}, \citenamefont {Duan},
  \citenamefont {Huang}, \citenamefont {He},\ and\ \citenamefont
  {Du}}]{Yin2025}%
  \BibitemOpen
  \bibfield  {author} {\bibinfo {author} {\bibfnamefont {P.}~\bibnamefont
  {Yin}}, \bibinfo {author} {\bibfnamefont {X.}~\bibnamefont {Xu}}, \bibinfo
  {author} {\bibfnamefont {K.}~\bibnamefont {Tian}}, \bibinfo {author}
  {\bibfnamefont {S.}~\bibnamefont {Lin}}, \bibinfo {author} {\bibfnamefont
  {Y.}~\bibnamefont {Sheng}}, \bibinfo {author} {\bibfnamefont
  {C.}~\bibnamefont {Yin}}, \bibinfo {author} {\bibfnamefont {D.}~\bibnamefont
  {Long}}, \bibinfo {author} {\bibfnamefont {C.-K.}\ \bibnamefont {Duan}},
  \bibinfo {author} {\bibfnamefont {P.}~\bibnamefont {Huang}}, \bibinfo
  {author} {\bibfnamefont {J.-h.}\ \bibnamefont {He}}, \ and\ \bibinfo {author}
  {\bibfnamefont {J.}~\bibnamefont {Du}},\ }\href {\doibase
  10.1038/s41550-024-02465-8} {\bibfield  {journal} {\bibinfo  {journal}
  {Nature Astronomy}\ }\textbf {\bibinfo {volume} {9}},\ \bibinfo {pages} {598}
  (\bibinfo {year} {2025})}\BibitemShut {NoStop}%
\bibitem [{\citenamefont {Clerk}\ \emph {et~al.}(2010)\citenamefont {Clerk},
  \citenamefont {Devoret}, \citenamefont {Girvin}, \citenamefont {Marquardt},\
  and\ \citenamefont {Schoelkopf}}]{Clerk2010}%
  \BibitemOpen
  \bibfield  {author} {\bibinfo {author} {\bibfnamefont {A.~A.}\ \bibnamefont
  {Clerk}}, \bibinfo {author} {\bibfnamefont {M.~H.}\ \bibnamefont {Devoret}},
  \bibinfo {author} {\bibfnamefont {S.~M.}\ \bibnamefont {Girvin}}, \bibinfo
  {author} {\bibfnamefont {F.}~\bibnamefont {Marquardt}}, \ and\ \bibinfo
  {author} {\bibfnamefont {R.~J.}\ \bibnamefont {Schoelkopf}},\ }\href
  {\doibase 10.1103/revmodphys.82.1155} {\bibfield  {journal} {\bibinfo
  {journal} {Reviews of Modern Physics}\ }\textbf {\bibinfo {volume} {82}},\
  \bibinfo {pages} {1155} (\bibinfo {year} {2010})}\BibitemShut {NoStop}%
\bibitem [{\citenamefont {Li}\ \emph {et~al.}(2023)\citenamefont {Li},
  \citenamefont {Lin}, \citenamefont {Zhang}, \citenamefont {Duan},
  \citenamefont {Huang},\ and\ \citenamefont {Du}}]{Li2023}%
  \BibitemOpen
  \bibfield  {author} {\bibinfo {author} {\bibfnamefont {R.}~\bibnamefont
  {Li}}, \bibinfo {author} {\bibfnamefont {S.}~\bibnamefont {Lin}}, \bibinfo
  {author} {\bibfnamefont {L.}~\bibnamefont {Zhang}}, \bibinfo {author}
  {\bibfnamefont {C.}~\bibnamefont {Duan}}, \bibinfo {author} {\bibfnamefont
  {P.}~\bibnamefont {Huang}}, \ and\ \bibinfo {author} {\bibfnamefont
  {J.}~\bibnamefont {Du}},\ }\href {\doibase 10.1088/0256-307x/40/6/069502}
  {\bibfield  {journal} {\bibinfo  {journal} {Chinese Physics Letters}\
  }\textbf {\bibinfo {volume} {40}},\ \bibinfo {pages} {069502} (\bibinfo
  {year} {2023})}\BibitemShut {NoStop}%
\bibitem [{\citenamefont {Kornack}\ \emph {et~al.}(2007)\citenamefont
  {Kornack}, \citenamefont {Smullin}, \citenamefont {Lee},\ and\ \citenamefont
  {Romalis}}]{Kornack2007}%
  \BibitemOpen
  \bibfield  {author} {\bibinfo {author} {\bibfnamefont {T.~W.}\ \bibnamefont
  {Kornack}}, \bibinfo {author} {\bibfnamefont {S.~J.}\ \bibnamefont
  {Smullin}}, \bibinfo {author} {\bibfnamefont {S.-K.}\ \bibnamefont {Lee}}, \
  and\ \bibinfo {author} {\bibfnamefont {M.~V.}\ \bibnamefont {Romalis}},\
  }\href {\doibase 10.1063/1.2737357} {\bibfield  {journal} {\bibinfo
  {journal} {Applied Physics Letters}\ }\textbf {\bibinfo {volume} {90}}
  (\bibinfo {year} {2007}),\ 10.1063/1.2737357}\BibitemShut {NoStop}%
\bibitem [{\citenamefont {O'Hare}(2020)}]{AxionLimits}%
  \BibitemOpen
  \bibfield  {author} {\bibinfo {author} {\bibfnamefont {C.}~\bibnamefont
  {O'Hare}},\ }\href {\doibase 10.5281/zenodo.3932430} {\enquote {\bibinfo
  {title} {cajohare/axionlimits: Axionlimits},}\ }\bibinfo {howpublished}
  {\url{https://cajohare.github.io/AxionLimits/}} (\bibinfo {year}
  {2020})\BibitemShut {NoStop}%
\bibitem [{\citenamefont {Tian}\ \emph {et~al.}(2025)\citenamefont {Tian},
  \citenamefont {Sheng}, \citenamefont {Li}, \citenamefont {Wang},
  \citenamefont {Yin}, \citenamefont {Lin}, \citenamefont {Long}, \citenamefont
  {Duan}, \citenamefont {Kong}, \citenamefont {Huang},\ and\ \citenamefont
  {Du}}]{Tian2025}%
  \BibitemOpen
  \bibfield  {author} {\bibinfo {author} {\bibfnamefont {K.}~\bibnamefont
  {Tian}}, \bibinfo {author} {\bibfnamefont {Y.}~\bibnamefont {Sheng}},
  \bibinfo {author} {\bibfnamefont {R.}~\bibnamefont {Li}}, \bibinfo {author}
  {\bibfnamefont {L.}~\bibnamefont {Wang}}, \bibinfo {author} {\bibfnamefont
  {P.}~\bibnamefont {Yin}}, \bibinfo {author} {\bibfnamefont {S.}~\bibnamefont
  {Lin}}, \bibinfo {author} {\bibfnamefont {D.}~\bibnamefont {Long}}, \bibinfo
  {author} {\bibfnamefont {C.-K.}\ \bibnamefont {Duan}}, \bibinfo {author}
  {\bibfnamefont {X.}~\bibnamefont {Kong}}, \bibinfo {author} {\bibfnamefont
  {P.}~\bibnamefont {Huang}}, \ and\ \bibinfo {author} {\bibfnamefont
  {J.}~\bibnamefont {Du}},\ }\href {\doibase 10.1103/physrevlett.134.111001}
  {\bibfield  {journal} {\bibinfo  {journal} {Physical Review Letters}\
  }\textbf {\bibinfo {volume} {134}},\ \bibinfo {pages} {111001} (\bibinfo
  {year} {2025})}\BibitemShut {NoStop}%
\end{thebibliography}%

\appendix 
\onecolumngrid
\bigskip
\hrule
\bigskip
\begin{center}
{\bf \large Supplemental Material}
\end{center}

\setcounter{equation}{0}
\setcounter{figure}{0}
\setcounter{table}{0}
\makeatletter
\renewcommand{\theequation}{S\arabic{equation}}
\renewcommand{\thefigure}{S\arabic{figure}}
\renewcommand{\thetable}{S\arabic{table}}

\section{Magnetostatics in Ferromagnetism}
\label{app:demag}
In the absence of free currents, two of the Maxwell equations are
\beq
\nabla\times {\bf H}=\nabla\cdot {\bf B}=0,
\eeq
The general solution is given by 
\beq
{\bf H}= -\nabla \Phi~.
\eeq 
Using $\bf B= H+M$ inside a ferromagnet, we have
\beq\label{eq:phi_in}
\nabla^2 \Phi_{\rm in}= \nabla\cdot{\bf M}~. 
\eeq
Outside, since $\bf M=0$, it reduces to Laplace equation
\beq\label{eq:phi_out}
\nabla^2 \Phi_{\rm out}= 0~.
\eeq
Furthermore, it is required that the tangential component of $H$ and normal component of $B$ are continuous across the boundary.

As an exercise, let's solve \cref{eq:phi_in,eq:phi_out} 
for some simple geometries involving uniformly magnetized ellipsoid, where $\nabla\cdot{\bf M}$ is zero everywhere except at the boundary. This is analogous to the situation as if magnetic monopoles formed on the surface and sourced a degmagnetizing field.
For a uniformly magnetized sphere with ${\bf M}= M_s\hat z$, one obtains 
$
{\bf H}_{\rm in}=-\frac13 M_s \hat z=-\frac13{\bf M}$. 
For a uniformly magnetized infinitely long cylinder with ${\bf M}= M_s\hat x$, one obtains $
{\bf H}_{\rm in}=-\frac12 M_s \hat x=-\frac12 {\bf M}$. 
In both cases, it is clear that ${\bf H}_{\rm in}$ are uniform and antiparallel to the magnetization with proportionality constants that do not depend on the type of ferromagnetic material. 

It turns out that for all uniformly magnetized ellipsoids, the demagnetizing field inside is 
\beq
{\bf H}_d =-D {\bf M}~,
\eeq
where $D$ is the demagnetizing tensor. If ${\bf M}$ is along one of the principal axes of the ellipsoid, $D$ can be diagonalized so that $D={\rm diag}(D_x,D_y,D_z)$ which satisfies ${\rm Tr} D=1$. Now for the special case of sphere, since it is completely spherically symmetric, $D_x=D_y=D_z=\frac13$. For the case of an infinitely long cylinder, since there are no surfaces in the $z$ direction, $D_z=0$, and due to the rotational symmetry in the plane, $D_x=D_y=\frac12$. 

One can imagine that in the case that the cylinder is very long with a finite height, $D_z\ll1$ but non zero.  This shape can be approximated by the case of a prolate spheroid, where the polar radius $c$ is much bigger than the equatorial radius $a$, like an egg shape. Using the notation
$
\gamma=c/a,~\xi =\frac{\sqrt{\gamma^2-1}}\gamma$,  
the demagnetizing factors are given by the following analytic expressions
\beq
D_z=\frac1{\gamma^2-1}\left[ \frac1{2\xi}\ln\left(\frac{1+\xi}{1-\xi}\right)-1\right],~D_x=\frac{1-D_z}2
\eeq
In the limit that $\gamma\gg1$, 
\beq
D_z\approx \frac{\ln \left(4 \gamma ^2\right)-2}{2 \gamma ^2}+\mathcal{O}(\gamma^{-4})
\eeq

According to the Brown-Morrish theorem, any uniformly magnetized ferromagnet behaves in the same way as an ellipsoid that has the same volume. This applies to a body that contains \emph{cavities} too.

\section{Numerical Calculations of Target Signal and Background Magnetic Noise}

\subsection{Target signal simulation}
\label{app:signal_simulation}
\begin{figure}[h]
    \centering
    \includegraphics[width=0.76\linewidth]{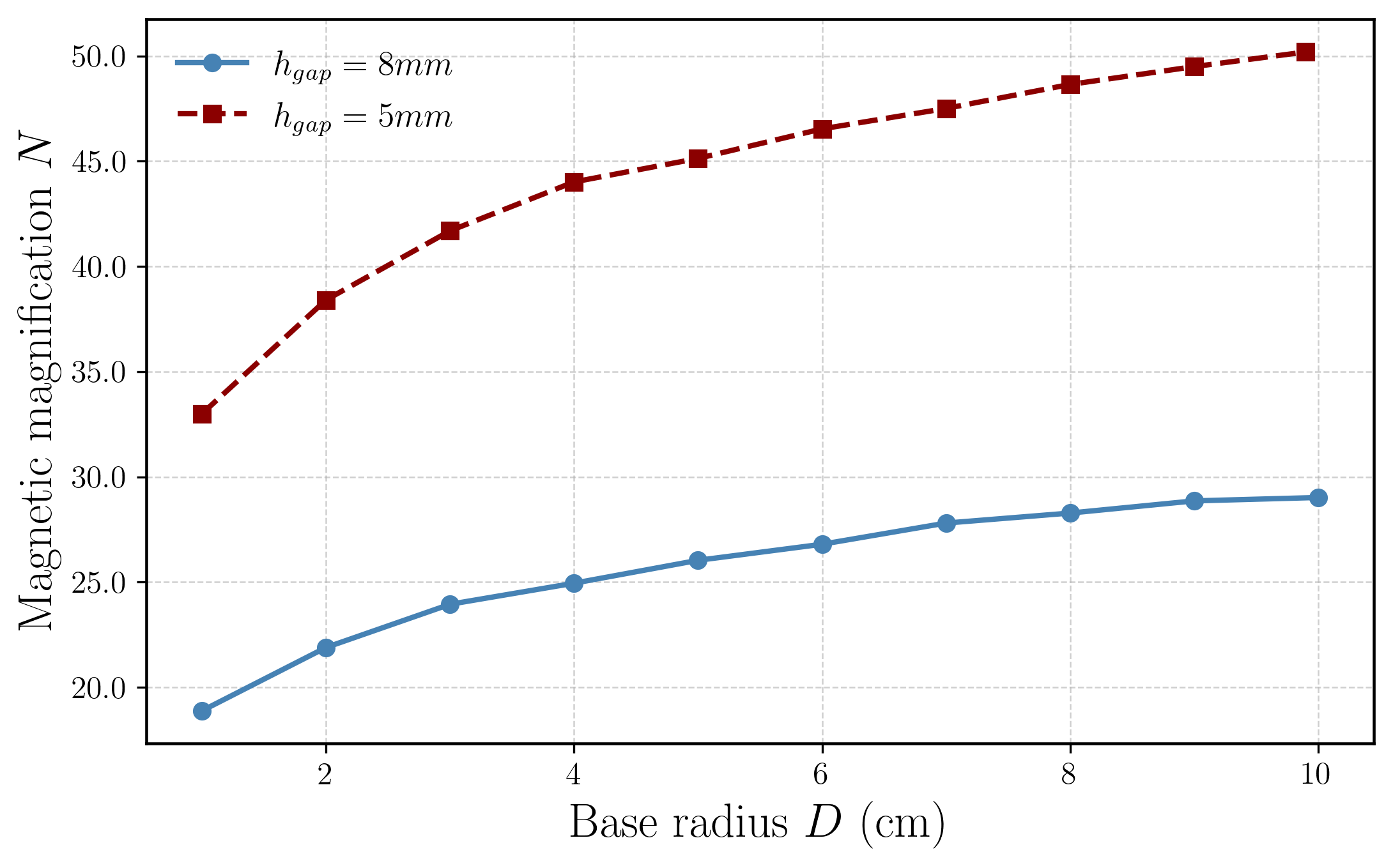}
    \caption{Simulated magnetic magnification $N$ as a function of base radius $D$ of the conical cores. Two cases are considered: gap size $h_{\rm gap} = 5$~mm and $h_{\rm gap} = 8$~mm. All other parameters follow the configuration detailed in the simulation setup.
}
    \label{fig:base_radius_comparison}
\end{figure}
We numerically solve the scalar magnetic potential equations \cref{eq:phi_ina,eq:phi_out}, subject to the boundary conditions defined in \cref{eq:phi_bca}. The simulation geometry consists of two conical soft magnetic cores, each with a tip radius of 3~mm and a height of 200~mm, aligned along the $y$-axis and separated by a gap of $h_{\rm gap} = 5$~mm or 8~mm. The entire structure is enclosed within a cubic domain with a side length of 5~m, filled with air or vacuum ($\mu_r = 1$), while the cones are assigned a relative permeability of $\mu_r = 3000$.

The problem is solved using finite element analysis (FEA). A locally refined mesh is employed, particularly in and around the cones. Upon convergence of the steady-state solution, the magnetic field components $B_y$ and $B_z$ are extracted on the $y$-$z$ plane at $x=0$.

Post-processing is conducted in Python. The exported field data are interpolated onto a uniform grid using cubic interpolation and then smoothed to reduce numerical noise while preserving the overall field structure. By exploiting the symmetry of the system, the data are reflected across the $z$-axis to reconstruct the full 2D field map. Streamlines are computed using a seeded integration approach, with streamlines initiated from regions of high field magnitude and along selected boundaries. The resulting magnetic field distribution and vector flow are visualized as a colormap overlaid with streamlines, as shown in Fig.~\ref{fig:field}.

To further investigate the geometric dependence of magnetic magnification, we perform a parametric sweep by varying the base radius $D$ of the conical cores from 1~cm to 10~cm, while keeping all other parameters fixed. The resulting magnification factor $N$ is shown in Fig.~\ref{fig:base_radius_comparison} for both $h_{\rm gap} = 5$~mm and $h_{\rm gap} = 8$~mm. In both cases, $N$ increases monotonically with $D$, with smaller gap sizes yielding substantially higher magnification due to more efficient flux concentration.

\subsection{Background magnetic field simulation}
\label{app:filed_bkg}
In our proposed setup, the ferrimagnetic concentrators are made of manganese-zinc (MnZn) ferrite. Manganese-zinc ferrites are a class of soft ferrimagnetic ceramic materials with the general formula (Mn,Zn)$\rm{Fe_{2}O_4}$. Their magnetic properties originate from the ferrimagnetic ordering of magnetic moments on the iron cations occupying the octahedral and tetrahedral sites within the spinel crystal lattice. A primary feature of MnZn ferrites is their high initial permeability, making them exceptionally effective for magnetic flux concentration at low frequencies. They also exhibit high saturation flux density and relatively low core losses, particularly in the low frequency range of several Hz to several hundred kHz. Compared to high permeability metal, MnZn ferrites have a much higher electrical resistivity, which enables an excellent suppression of the eddy current and the corresponding Johnson noise currents.

\begin{figure}[h]
    \centering
    \includegraphics[width=1.05\linewidth]{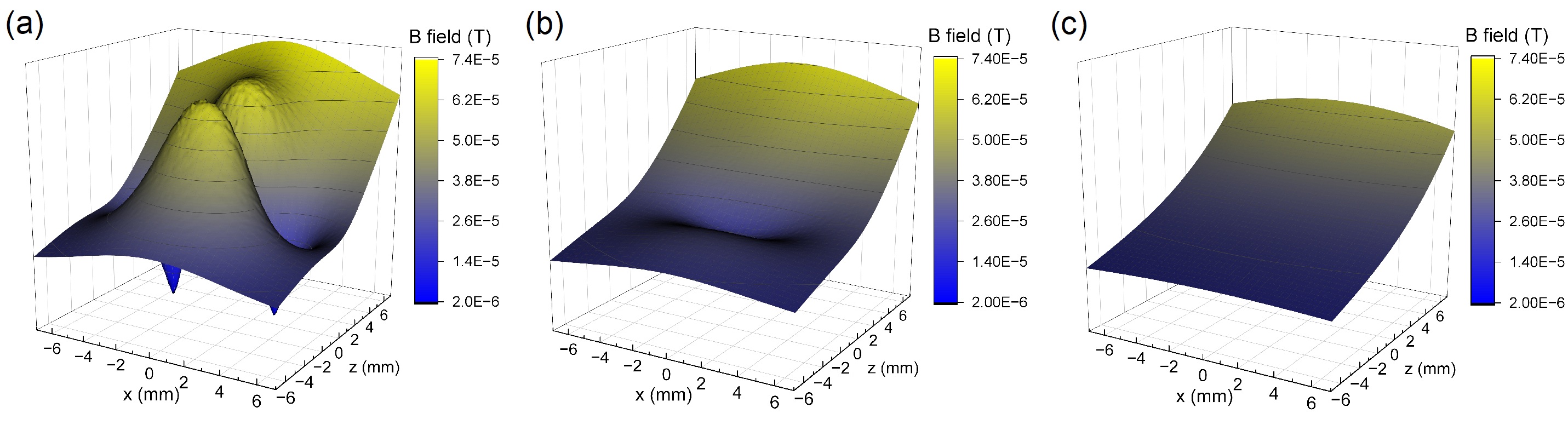}
    \caption{The background magnetic field generated by the lifting magnet and the levitated ferromagnet. Panels (a)-(c) display the results within the domain $-7~\rm{mm}<x, z<7 ~\rm{mm}$ with $y$-axis positions of $y=2.5, 5$, and 10 mm, respectively.  }
    \label{fig:b_bg}
\end{figure}

At low frequency and weak magnetic field, a complex relative permeability of $\mu_{\rm r}'=3000, \mu_{\rm r}''=0.2$ is achievable. To maintain high magnetic permeability, the background magnetic field around the tips of the flux concentrators, which is generated by the lifting magnet and the levitated ferromagnet, must be controlled below 0.1 mT. In our proposed setup, the typical geometry of lifting magnet and the levitated ferromagnet is chosen to have the following properties: the lifting magnet is a cylinder of diameter 4 mm and length 2.5 mm; the levitated ferromagnet is a cylinder of diameter 60 $\mu$m and length 5 mm; and the distance between them is 35 mm. The numerical simulation results for the background magnetic field in the $x-z$ plane are presented in Fig.~\ref{fig:b_bg}. For the three scenarios studied, the average magnetic field within the region of $-7~\rm{mm}<x, z<7~\rm{mm}$ on the $x-z$ plane is calculated to be 0.042 mT, 0.035 mT, and 0.032 mT, respectively. Thus we confirm that high magnetic permeability can be achieved.

\subsection{Background magnetic noise simulation}
To calculate the magnetic noise generated in the ferromagnetic concentrators due to magnetic hysteresis loss, we approximate the levitated ferromagnet as an oscillating current flowing in an excitation coil with an equivalent magnetic moment $m=i(t) S=MV\theta(t)$, located at the same point~\cite{Kornack2007}. Here, $i(t)$ is the current and $S$ is the area of the coil. The magnetic noise is calculated by \cref{eq:Sfer-theory}. The average dissipation power $P=\int_V \frac{1}{2}\omega \mu_{\rm r}''\mu_0 H^2 \rm {d} V$ is calculated by finite element simulations. 

\begin{figure}[h]
    \centering
    \includegraphics[width=0.8\linewidth]{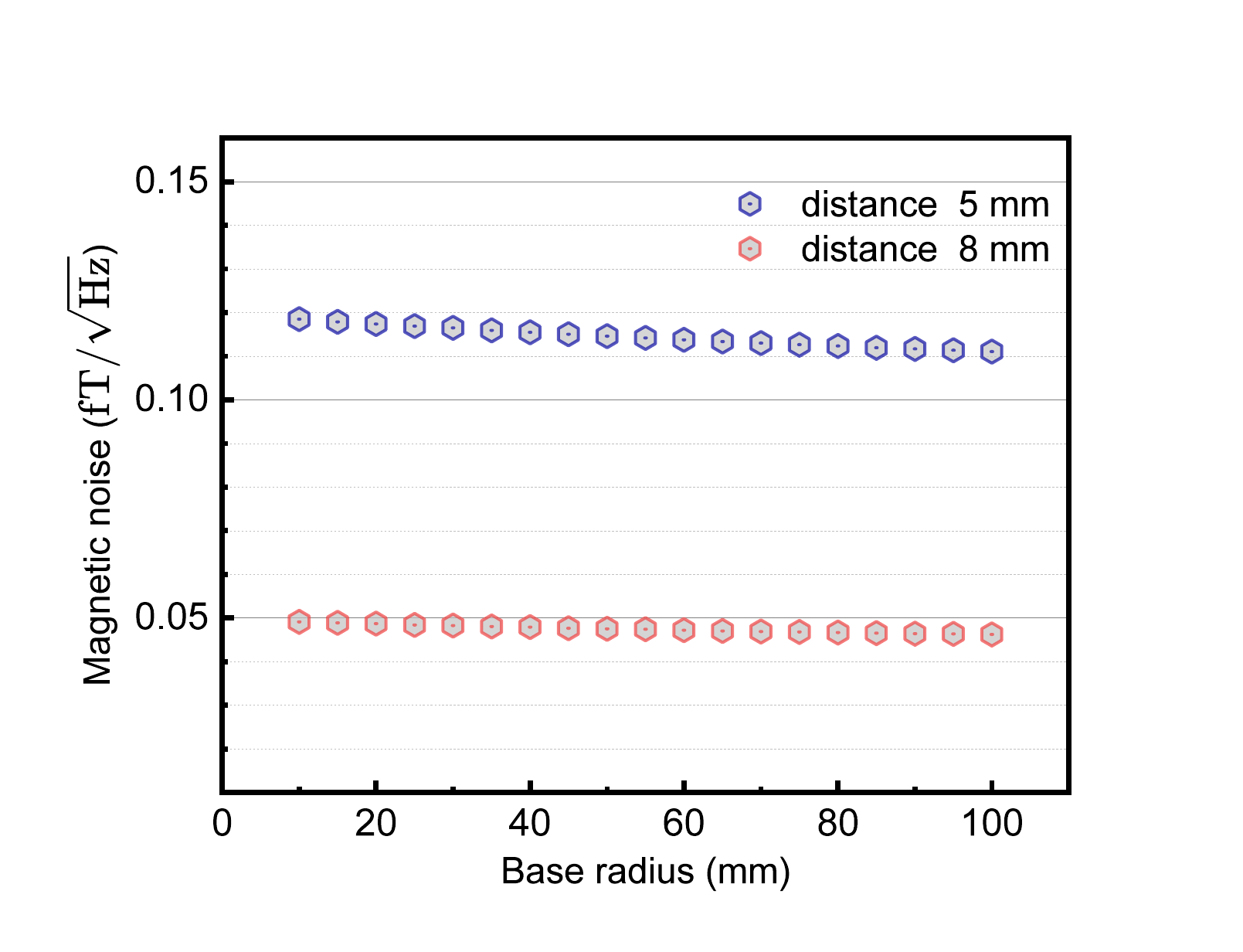}
    \caption{Magnetic noise applied on the levitated ferromagnet (\cref{fig:setup}) by the ferrimagnetic flux concentrator as a function of the base radius of each concentrator, assuming a temperature of 3 K and at a frequency of 200 Hz. It is assumed that each cone-shaped concentrator has a tip radius of 3 mm and a height of 200 mm. And we choose the complex relative permeability as $\mu_{\rm r}'=3000, \mu_{\rm r}''=0.2$.  We compare two choices of gaps between the tips of the concentrators: 5 mm (blue) and 8 mm (red). }
    \label{fig:mag_noise}
\end{figure}

In the FEA, we take same geometry as in \cref{app:signal_simulation} but replace the levitated ferromagnet with a coil. We choose the complex relative permeability of the ferromagnetic flux concentrators as $\mu_{\rm r}'=3000, \mu_{\rm r}''=0.2$. And we obtain the numerical solution in a frequency domain, with the drive current oscillating at frequency 200 Hz. The results of magnetic noise is given in Fig.~\ref{fig:mag_noise}.

\label{app:bkg}

\end{document}